\documentclass{aa}
\usepackage{graphicx}
\begin{document}
\title{Data reduction methods for single-mode optical interferometry}
\subtitle{Application to the VLTI two-telescopes beam combiner VINCI}
   \author{	P. Kervella\inst{1,3},
		D. S\'egransan\inst{2}
		\and V. Coud\'e du Foresto\inst{1}}
 		\offprints{P. Kervella}

   \institute{
	LESIA, UMR 8109, Observatoire de Paris-Meudon, 5 pl. Jules Janssen, F-92195 Meudon Cedex, France\\
              \email{Pierre.Kervella@obspm.fr}
         \and
         Observatoire de Gen\`eve, 51 ch. des Maillettes, CH-1290 Sauverny, Switzerland
	\and
   	European Southern Observatory,
              Alonso de Cordova 3107, Casilla 19001, Santiago 19, Chile
             }
 \date{Received 23 October 2003 ; Accepted 16 June 2004}

\authorrunning{P. Kervella et al.}
\titlerunning{Data reduction methods for optical interferometry}

 \abstract{The interferometric data processing methods that we describe
in this paper use a number of innovative techniques. In particular, the
implementation of the wavelet transform allows us to obtain a
good immunity of the fringe processing to false detections and
large amplitude perturbations by the atmospheric piston effect,
through a careful, automated selection of the interferograms.
To demonstrate the data reduction procedure,
we describe the processing and calibration of a sample of stellar data from
the VINCI beam combiner. Starting from the raw data, we
derive the angular diameter of the dwarf star $\alpha$\,Cen\,A.
Although these methods have been developed specifically for VINCI,
they are easily applicable to other single-mode beam combiners, and
to spectrally dispersed fringes.
}
\maketitle
%

\section{Introduction}

Although interferometric techniques are now used routinely around
the world, the processing of  interferometric data is still the subject of active
research. In particular, the corruption of the interferometric
fringes by the turbulent atmosphere is currently the most critical limitation
to the precision of the ground-based interferometric measurements.

Installed at the Very Large Telescope Interferometer (VLTI),
the VINCI instrument coherently combines the infrared light
coming from two telescopes in the infrared $H$ and $K$ bands.
The first fringes were obtained in March 2001 with the VLTI Test Siderostats, and
in October 2001 with the 8m Unit Telescopes (UTs).
To reduce the large quantity of data produced by this
instrument, we have developed innovative interferometric
data analysis methods, using in particular the wavelet transform.
We have appplied them successfully to a broad range
of interferometric observations obtained with very different
configurations of the VLTI (0.35\,m siderostats, 8\,m Unit Telescopes,
16\,m to 140\,m baselines, $K$ and $H$ band observations).

Since the first fringes of VINCI, more than 800 nights of observations have been conducted
with this instrument. This allowed us to record a large number of individual star observations,
under extremely different atmospheric and instrumental conditions.
The data processing methods that are described in the present paper
were successfully applied to all these configurations, and
consistently provided reliable and precise results.
This gives us good confidence that they are efficient and robust, and can be
generalized to other interferometric instruments.

\begin{figure}[t]
\centering
\includegraphics[bb=0 10 520 288, width=16cm,angle=-90]{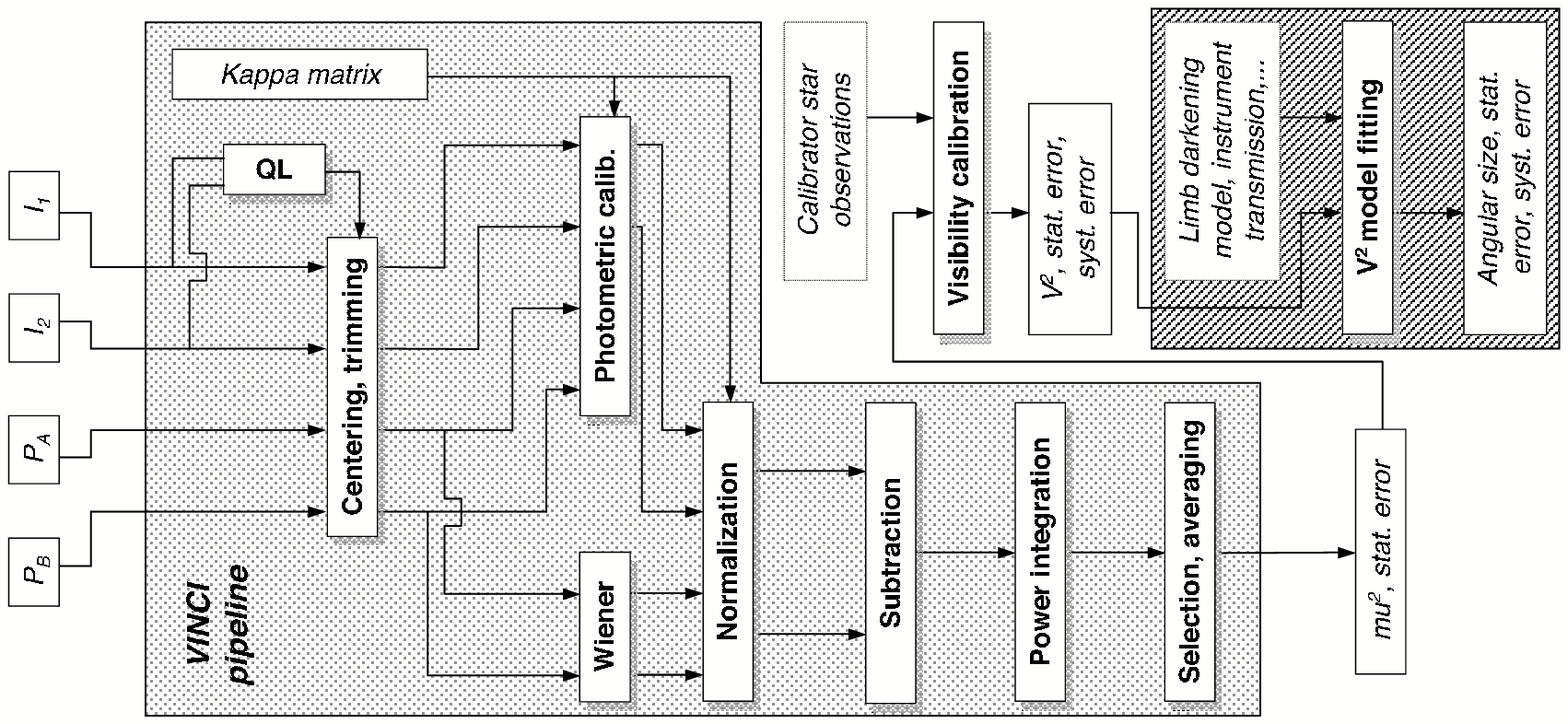}
\caption{Overview of the VINCI data analysis work flow.
The shaded area delimits the processing executed
automatically by the instrument data pipeline. The hatched
area (lower right) covers the astrophysical interpretation of the
measured visibility, not adressed in the present paper.}
\label{diagram}
\end{figure}

Our goal in this paper is to give a step by step description of the processing
of the VINCI data, from the raw data to the calibrated visibility.
To illustrate this processing, we selected from the commissioning data
a series of observations of a bright star and its calibrator,
$\alpha$\,Cen\,A and $\theta$\,Cen respectively (Sect.\,\ref{sample_data}).
A complete overview of the data analysis work flow is presented in Fig.~\ref{diagram}.
It can be used as a reference to follow the logical progression of this paper.
The photometric calibration of the interferograms is described in
Sect.\,\ref{preliminary_steps} and \ref{photometric_calibration}. The criteria for the selection of the
interferograms are detailed in Sect.\,\ref{selection_interf}, and the computation of the visibility
values and associated errors is given in Sect.\,\ref{power_integration}.
A number of quality controls is applied to the reduced data; they are described in
Sect.\,\ref{quality_control}. The calibration of the visibility is illustrated in Sect.\,\ref{visibility_calibration}.
We demonstrate in particular the computation of the statistical and systematic errors on the visibility values.

\section{Instrument description}
\subsection{The VLT Interferometer and VINCI}

The Very Large Telescope Interferometer
(VLTI, Glindemann et al.~\cite{glindemann};
Glindemann et al.~\cite{glindemann03a};
Glindemann et al.~\cite{glindemann03b};
Sch\"oller et al.~\cite{schoeller03}) has been operated by the
European Southern Observatory on
top of the Cerro Paranal, in Northern Chile since March 2001.
In its current state of completion, the light coming from two telescopes
can be combined coherently in VINCI, the VLT Interferometer
Commissioning Instrument (Fig.~\ref{VINCI_photo}),
or in the MIDI instrument (Leinert et al. \cite{leinert00}).
In December 2002, MIDI obtained its first fringes at
$\lambda = 8.7\,\mu$m between the two 8m Unit Telescopes
Antu (UT1) and Melipal (UT3).
Another instrument, AMBER (Petrov et al.~\cite{petrov00}) will soon
allow the simultaneous recombination of three telescope beams
(its first observations are scheduled for 2004).

\begin{figure}[t]
\centering
\caption{View of the VINCI instrument installed in the VLTI
interferometric laboratory. The MONA beam combiner is the
visible above the center of the image (white box), with its
optical fiber inputs and outputs. The beams coming from the VLTI
Delay Lines enter the optical table from the bottom of the picture.
{\it Figure not available on the astro-ph version.}}
\label{VINCI_photo}
\end{figure}

\begin{figure}[t]
\centering
\caption{Principle of the VINCI instrument. {\it Figure not available on the astro-ph version.}}
\label{VINCI_starintf}
\end{figure}

A detailed description of the VINCI instrument, including its hardware and software
design, can be found in Kervella et al.~(\cite{kervella00}).
Fig.\,\ref{VINCI_starintf} shows the setup of VINCI. The two beams enter the instrument
from the bottom of the figure, after having been delayed by two optical delay lines (Derie~\cite{derie00}).
Once the stellar light from the two telescopes has been injected into the optical fibers (injections A and B),
it is recombined in the MONA triple coupler.
VINCI is based on the same principle as the FLUOR instrument
(Coud\'e du Foresto et al.~\cite{coude98}), and recombines the light through single-mode fluoride
glass optical fibers that are optimized for $K$ band operation ($\lambda=2.0-2.4\ \mu$m).
It uses in general a regular $K$ band filter, but can also observe
in the $H$ band ($\lambda=1.4-1.8\ \mu$m) using an integrated optics beam combiner
(Berger et al.~\cite{berger01}). The first observations with this new generation coupler
 installed at the VLTI focus have given promising results
(Kervella et al.~\cite{kervella03a}; Kern et al.~\cite{kern03}).
\subsection{Beam combination}

The central element of VINCI is its optical correlator (MONA), based on single-mode
fluoride glass fibers and couplers.
It was designed and built specifically for VINCI by the company Le Verre Fluor\'e (France).
The waveguides are used to filter out the spatial modes of the atmospheric turbulence.
In the couplers, the fiber cores are brought very close to each other (a few $\mu$m) and
the two electric fields interfere directly with each other
by evanescent coupling of the electromagnetic waves.
Motorized polarization controllers allow the matching of the beam polarizations, in order to obtain
the best possible interferometric transfer function.

The general principle of the MONA box is shown in Fig.\,\ref{MONA_principle}.
MONA contains three couplers: two side couplers (that provide two photometric outputs
$P_A$ and $P_B$ to monitor the efficiency of the stellar light injection in the optical fibers) and
a central coupler that is used for the beam combination.  The latter provides
two complementary interferometric outputs $I_1$ and $I_2$.
The four output fibers are eventually arranged on a 125\,$\mu$m
square and imaged onto an infrared camera (LISA), built around a HAWAII detector.
Only four small windows, of one pixel each, are read from the detector to
increase the frame frequency and reduce the readout noise.

The Optical Path Difference (OPD) between the two beams is modulated by a mirror
mounted on a piezo translator.
This modulation allows one to sweep through the interference fringes (at zero OPD), that
appear as temporal modulations of the $I_1$ and $I_2$ intensities on the detector.
While the OPD is scanned, the four output signals are sampled at a few kHz.
The four resulting time sequence signals (two photometric and
two interferometric) are then available for processing. The interferogram acquisition
rate can be set between 0.1 Hz (faint objects) and 20 Hz (bright targets).

\begin{figure}[t]
\centering
\caption{Principle of the MONA beam combiner. {\it Figure not available on the astro-ph version.}}
\label{MONA_principle}
\end{figure}
\subsection{Coherencing}

During the observations, a simple fringe packet centroiding algorithm is applied
in near real-time to the raw data. The fringe packet center is localized with a
precision of about one fringe (2 $\mu$m) after each scan and the resulting error is fed back
to the VLTI delay lines as an OPD offset. This capability, called {\it fringe coherencing}, ensures
that the residual OPD is less than a coherence length despite possible instrumental
drifts. Still, the correction rate (once per scan, i.e. a few Hz) is too slow to remove
the differential piston mode of the turbulence. A fringe tracking unit is anticipated for the
VLTI (FINITO, Gai et al.~\cite{gai03}) that will remove the differential piston
and stabilize the interference pattern at the sub-fringe level (fringe cophasing), thus enabling
longer integration times for the scientific instrument.
\section{The selected sample data sets \label{sample_data}}
\subsection{Targets}

To illustrate the processing of the VINCI data on representative files, we have chosen
two series of interferograms obtained respectively on a calibrator star, $\theta$\,Cen, and a target of
scientific interest, $\alpha$\,Cen\,A, on the intermediate length E0-G1 baseline (66\,m ground length).
$\theta$\,Cen was chosen from the Cohen et al.~(\cite{cohen99}) catalogue. These authors
compiled a grid of calibrator stars whose angular diameter is typically known with a relative
precision better than 1\%. Bord\'e et al.~(\cite{borde02}) recently revised this catalogue
specifically for its application to long baseline interferometry.

The observations of $\alpha$\,Cen\,A and $\theta$\,Cen discussed here
were carried out with the two 0.35\,m test siderostats of the VLTI.
Both stars are bright, but $\theta$\,Cen is significantly smaller
than $\alpha$\,Cen\,A, therefore its visibility is higher.
The relevant properties of $\theta$\,Cen and $\alpha$\,Cen
are reported in Table~\ref{params}.
The angular diameter of $\alpha$\,Cen\,A was
measured for the first time by Kervella et al.~(\cite{kervella03b}),
based on a series of observations that include the two data sets
discussed here.

The file names and characteristics of the two selected data sets are
given in Table\,\ref{data_sets}, to allow the interested reader to
retrieve them from the ESO Archive ({\it http://archive.eso.org/}).

   \begin{table}
      \caption[]{Relevant parameters of $\theta$\,Cen and $\alpha$\,Cen\,A.}
	\label{params}	
\begin{tabular}{lcc}
& $\theta$\,Cen & $\alpha$\,Cen\,A\\
& \object{HD 123139} & \object{HD 128620}  \\
\hline
\noalign{\smallskip}
$m_\mathrm{V}$ & 2.1 & -0.0\\
$m_\mathrm{K}$ & -0.1 & -1.5 \\
Spectral Type & K0IIIb & G2V\\
$\rm T_{\mathrm{eff}}$ (K)$^{\mathrm{a}}$ & 4980 & 5790 \\
$\log g^{\mathrm{a}}$ & 2.75 & 4.32 \\
$[{\rm Fe}/{\rm H}]^{\mathrm{a}}$ & 0.03 & 0.20 \\
$\theta_{\rm {UD}}$ (mas)$^{\mathrm{b}}$ & 5.305 $\pm$ 0.020 & 8.314 $\pm$ 0.016\\
\noalign{\smallskip}
\hline
\end{tabular}
\begin{list}{}{}
\item[$^{\mathrm{a}}$] $\rm T_{\mathrm{eff}}$, $\log g$ and $[{\rm Fe}/{\rm H}]$ from
Cayrel de Strobel et al.~(\cite{cayrel01}) and Morel et al.~(\cite{morel00}),
respectively for $\theta$\,Cen and $\alpha$\,Cen\,A.
\item[$^{\mathrm{b}}$] Measured diameters from Kervella et al.~(\cite{kervella03b}).
\end{list}
\end{table}
\subsection{Acquisition parameters and data structure\label{data_struct}}

\begin{figure*}[t]
\centering
\includegraphics[bb=0 0 720 300, width=17cm]{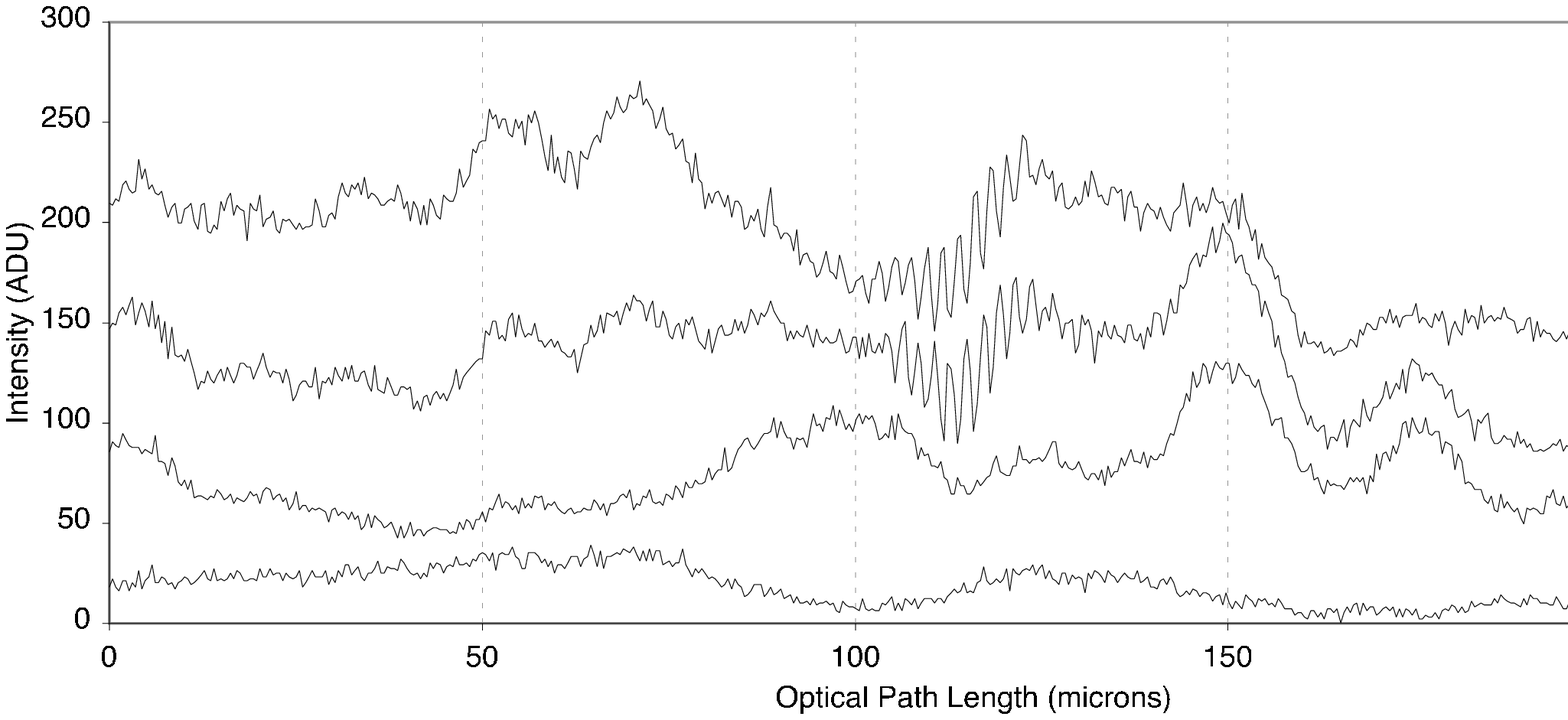}
\caption{The raw signals $I_1$, $I_2$, $P_A$ and $P_B$, for one interferogram obtained on $\theta$\,Cen.
The original signals have been translated vertically respectively by +90, +40, +40 and 0 ADUs for clarity.}
\label{Raw_signals}
\end{figure*}

Following the standard procedure used with VINCI, a series of 500 interferograms
was obtained on each object. The two data sets were taken on July 15, 2002, starting at
UT times 01:32:32 for $\theta$\,Cen, and 02:33:09 for $\alpha$\,Cen\,A.
The piezo mirror scanning speed was set to 650\,$\mu$m/s, giving
a fringe frequency of 297\,Hz.
This intermediate speed is used commonly for the operation
of VINCI with the VLTI Test Siderostats.
The LISA camera frequency was set to 1.5\,kHz in order
to obtain a sampling of 5 points per fringe.
The choice of the scanning speed (hence the sampling rate of the
camera) is the result of a compromise between the photometric SNR
and the phase perturbations of the atmosphere (dominant at low scanning speed).
The VINCI instrument allows one to scan up to a fringe frequency of 680\,Hz (camera frequency of 3.4\,kHz).
This extreme speed is useful in the case of observations with the 8\,m Unit Telescopes
to reduce the influence of the photometric fluctuations on the interference fringes (multi-speckle regime).
Fig.\,\ref{Raw_signals} shows the raw signals of one interferogram obtained on $\theta$\,Cen.
This is the second scan in the series of 500, and it is of average quality in terms of injected
flux stability. The photometric fluctuations are clearly visible in all four channels,
while the interference fringes are located close to the center of the scan.
The fringes are naturally in phase opposition between the two
channels $I_1$ and $I_2$.

Each star observation consists of four files (batches), that
each contain a series of acquisitions (scans) of the four signals coming out of MONA,
with four different configurations of the instrument.
The first three batches are used for the calibration of the
camera background and instrument transmission, and usually contain 100 scans.
The fourth batch contains the fringes. They are recorded in the following sequence:
\begin{enumerate}
  \item {\it Off-source}: the two injection parabolas of VINCI are displaced in order to remove the star image
from the single-mode fiber head.
  \item {\it Beam A}: the A injection parabola is brought back to the position where
the star light is injected in the optical fiber, while B is still off.
  \item {\it Beam B}: symmetrically, only the beam B is injected in the MONA beam combiner,
while A is off. The {\it Beam A} and {\it Beam B} sequences are used in the computation of the
$\kappa$ matrix (Sect.\,\ref{kappa_matrix_sect}).
  \item {\it On-source}: both beams are injected into MONA, and interference can occur.
This last series usually contains 500 scans and is used to compute the squared coherence
factor $\mu^2$  (the interferometric observable, defined in Sect.~\ref{quality_interf}).
\end{enumerate}

   \begin{table}
      \caption[]{Sample data sets. $N$ is the number of scans.}
	\label{data_sets}	
\begin{tabular}{lccc}
File name & Target & Type & $N$ \\
\hline
\noalign{\smallskip}
VINCI.2002-07-15T01:30:12.042 & $\theta$\,Cen & Off & 100\\
VINCI.2002-07-15T01:30:53.273 & $\theta$\,Cen & A & 100\\
VINCI.2002-07-15T01:31:36.505 & $\theta$\,Cen & B & 100\\
VINCI.2002-07-15T01:32:31.645 & $\theta$\,Cen & On & 500\\
\noalign{\smallskip}
VINCI.2002-07-15T02:30:49.318 & $\alpha$\,Cen\,A & Off & 100\\
VINCI.2002-07-15T02:31:30.046 & $\alpha$\,Cen\,A & A & 100\\
VINCI.2002-07-15T02:32:10.840 & $\alpha$\,Cen\,A & B & 100\\
VINCI.2002-07-15T02:33:08.661 & $\alpha$\,Cen\,A & On & 500\\
\noalign{\smallskip}
\hline
\end{tabular}
\end{table}
 
\section{Preliminary steps \label{preliminary_steps}}
\subsection{Computation of the $\kappa$ matrix \label{kappa_matrix_sect}}

To properly calibrate the photometric fluctuations of the interferometric signals I$_1$ and I$_2$  using the
two photometric outputs P$_A$ and P$_B$, it is necessary to know precisely the coefficients linking the
intensities of these four outputs. The relationships between the four channels can be
approximated, within a very good precision (Coud\'e du Foresto et al.~\cite{cdf97}), by the
following expressions:
\begin{eqnarray}
{I}_1 = \kappa_{1,A}\ {P}_A + \kappa_{1,B}\ {P}_B\\
{I}_2 = \kappa_{2,A}\ {P}_A + \kappa_{2,B}\ {P}_B
\end{eqnarray}
The $\kappa$ coefficients correspond to the differential gains between the four
channels of VINCI. They include the coupling ratios of the MONA box, the coupling
efficiency of each fiber to the physical pixels of the infrared camera, and the differential
quantum efficiency between these pixels.
Due to the chromatic transmission of the couplers,
the color of the observed object plays a role in the $\kappa$ coefficients,
and they also tend to evolve due to the slow motion of the
fiber spots on the LISA detector.
It is therefore necessary to measure these coefficients (the $\kappa$ matrix)
immediately before each star observation.
Each pair of $\kappa$ coefficients is computed simultaneously using a
classical $\chi^2$ minimization algorithm with two variable parameters.
The errors on the estimation of the $\kappa$ coefficients are derived
from the residual dispersion of the measurement points around the linear model.

Table~\ref{kappa_values} gives the $\kappa$ coefficients derived for the
$\theta$\,Cen and $\alpha$\,Cen observations.
The small differences between the $\kappa$ values for the two stars may come from the
slightly different colors of these objects, or from a small variation of the alignment of the output
spots on the LISA infrared camera pixels between the two observations
(they are separated in time by $\Delta t \simeq 1$\,h).

Ideally, the $\kappa$ coefficients should be balanced between the four outputs
in order to maximize the efficiency of the interference, and simultaneously
to give high SNR photometric signals for the calibration of the interferograms.
The observed inbalance (that can reach up to a factor 5 in the selected sample batches)
is due to the fact that the unexpectedly fast aging of the three optical couplers in the
MONA box has increased significantly their sensitivity to temperature.
This effect cannot be corrected on the coupler itself, and causes a slow
(timescale of months) but large amplitude evolution of the $\kappa$ matrix.
Due to the very different time scales of these variations (months) and of the scientific
observations (hours), this sensitivity is expected to have no significant impact on the
observations other than a uniform and moderate reduction of the quality
of the LISA signals.

\begin{table}
\caption[]{$\kappa$ coefficients measured for $\theta$\,Cen and $\alpha$\,Cen\,A.}
\label{kappa_values}	
\begin{tabular}{lcc}
& $\theta$\,Cen & $\alpha$\,Cen\,A\\
\hline
\noalign{\smallskip}
$\kappa_{1A}$ & $0.7569 \pm 0.0061$ & $0.7576 \pm 0.0037$ \\
$\kappa_{1B}$ & $4.1231 \pm 0.0160$ & $4.1877 \pm 0.0120$ \\
$\kappa_{2A}$ & $1.3089 \pm 0.0054$ & $1.2790 \pm 0.0044$ \\
$\kappa_{2B}$ & $2.4855 \pm 0.0143$ & $2.4735 \pm 0.0061$ \\
\noalign{\smallskip}
\hline
\end{tabular}
\end{table}

\subsection{Fringe localization \label{fringe_localization}}

The first step of the processing is to trim the long interferogram to restrict it to a shorter
segment, where the fringes are centered. The detection of the
fringes is then achieved with the {\it Quicklook} signal QL, that is computed using the simple formula:
\begin{equation}
{QL} = {I}_1 - a\ {I}_2
\end{equation}
where $a$ is given by
\begin{equation} 
a = \frac{\sum_{i = 1}^{N}{{I}_1(i)\ {I}_2(i)}}{\sum_{i=1}^{N}{{I}_1^2(i)}}
\end{equation}
where $N$ is the number of samples of the raw signals $I_1$ and $I_2$.
This operation attenuates the photometric fluctuations and increases the SNR of the fringes.
Once the fringes are localized precisely by detecting the maximum of the
wavelet power spectral density (WPSD) of the QL, the four raw signals
are trimmed accordingly.
If the fringes have been found too close to the edge of the interferogram,
the scan is discarded to avoid any bias of the data. In addition, if a large amplitude
jump of the position of the fringe packet is detected between two consecutive scans
(more than 20\,$\mu$m), a strong piston effect is suspected and the scan is
rejected (see also Sect.\,\ref{selection_interf}).
\section{Photometric calibration \label{photometric_calibration}}
\subsection{General principle}

\begin{figure}[t]
\centering
\includegraphics[bb=0 0 360 288, width=8.5cm]{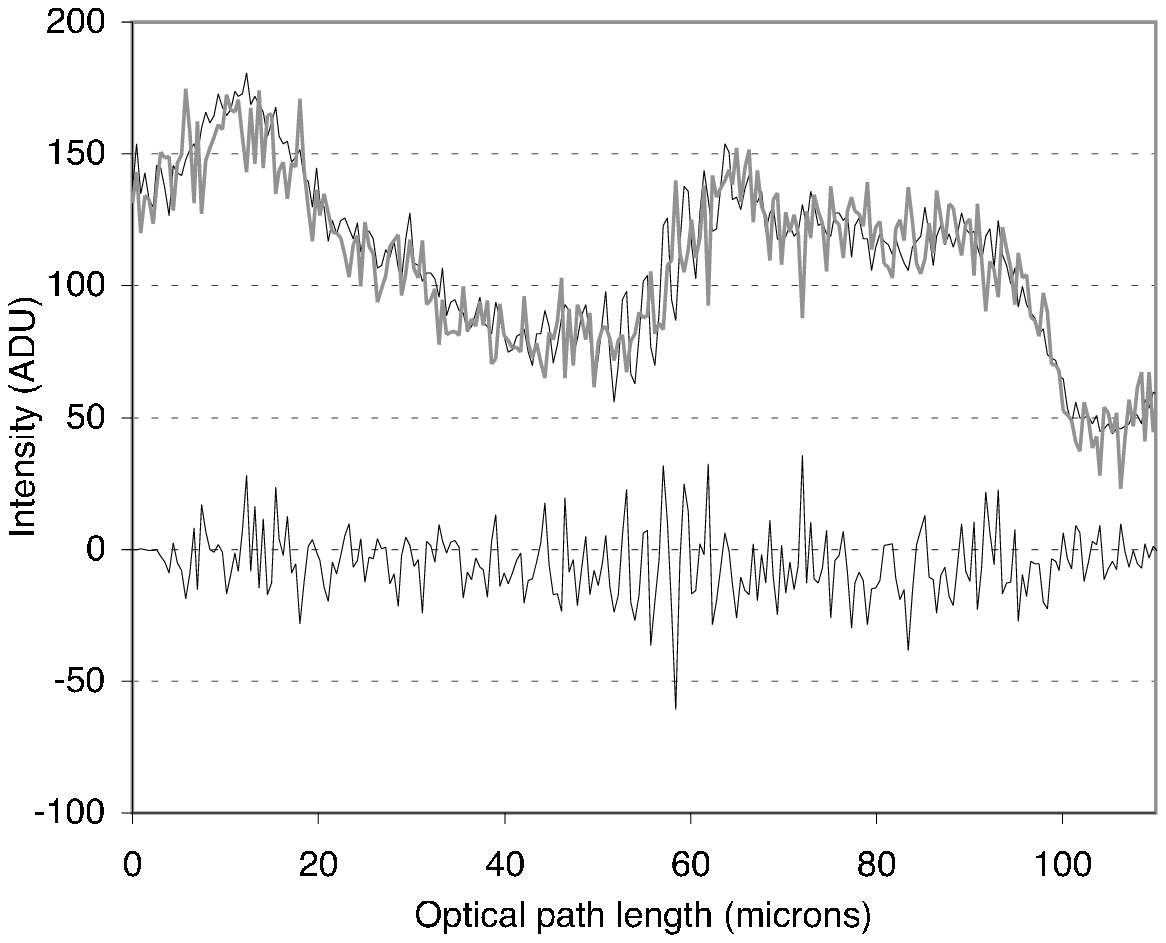}
\caption{Photometric calibration of the $I_1$ signal. The raw $I_1$ signal is the black line in
the upper part of the plot, the photometric calibration signal $\kappa_{1A} P_A - \kappa_{1B} P_B$ is
the superimposed grey line, and the result of the subtraction is the lower curve.}
\label{I1calib}
\end{figure}

The photometric calibration of the interferograms produced by VINCI is achieved
using the two photometric control signals P$_A$ and P$_B$ and
the $\kappa$-matrix. The calibration is computed separately for the I$_1$ and I$_2$ channels
using the following formulae (see Coud\'e du Foresto et al.~\cite{cdf97} for their derivation):
\begin{equation}\label{calib_I1}
I_{1\,cal} = \frac{1}{2\ \sqrt{\kappa_{1,A}\ \kappa_{1,B}}}
\frac{I_1 - \kappa_{1,A} P_A - \kappa_{1,B} P_B}{[\sqrt{P_A\ P_B}]_{\mathrm{Wiener}}}
\end{equation}
\begin{equation}\label{calib_I2}
I_{2\,cal} = \frac{1}{2\ \sqrt{\kappa_{2,A}\ \kappa_{2,B}}}
\frac{I_2 - \kappa_{2,A} P_A - \kappa_{2,B} P_B}{[\sqrt{P_A\ P_B}]_{\mathrm{Wiener}}}
\end{equation}
The subscript ``Wiener" designates optimally filtered signals (Sect.\,\ref{wiener}).
This process allows one first to subtract the photometric fluctuations that were introduced
in the interferometric channels by the turbulent atmosphere (calibration),
and then to normalize the resulting signals to the geometrical mean of the two photometric
channels $P = \sqrt{P_A\ P_B}$.
As an example of calibration, the subtraction of
$\kappa_{1,A} P_A + \kappa_{1,B} P_B$ from the original $I_1$ is
presented in Fig.\,\ref{I1calib}.
The Wiener filtering of $P$, essential to avoid numerical instabilities,
is described in the next paragraph (Sect.\,\ref{wiener}).
After the normalization, $I_{1\ {\rm cal}}$ and $I_{2\ {\rm cal}}$
are apodized at their extremities, to prevent any 
edge effect during the numerical wavelet transform.
\subsection{Low pass filtering of photometric signals}\label{wiener}

The normalization by the $P$ signal is a critical step of the calibration.
If $P$ presents too low values (``zero crossing''),
the divisions of Eq.\,\ref{calib_I1} and \ref{calib_I2} will amplify the noise of the numerator.
This is the reason why the $P_A$ and $P_B$ signals have to be filtered, to improve their
SNRs. This is achieved using Wiener filters, that allow one to optimally filter the raw signal
and to reject the detector noise.
They are computed from the average power spectral density (PSD) of the photometric fluctuations
of $P_A$ and $P_B$ using only the on-source spectra.
We use the classical definition of the Wiener filter $W_x$ as computed from the
signal $P_x$ and the noise $N_x$
(with $x=A$ or $B$, the Fourier transform being represented by the $\sim$ notation):
\begin{equation}
W_x = \frac{\widetilde{P_x}^2-\widetilde{N_x}^2}{\widetilde{P_x}^2}
\end{equation}

\begin{figure}[t]
\centering
\includegraphics[bb=0 0 360 288, width=8.5cm]{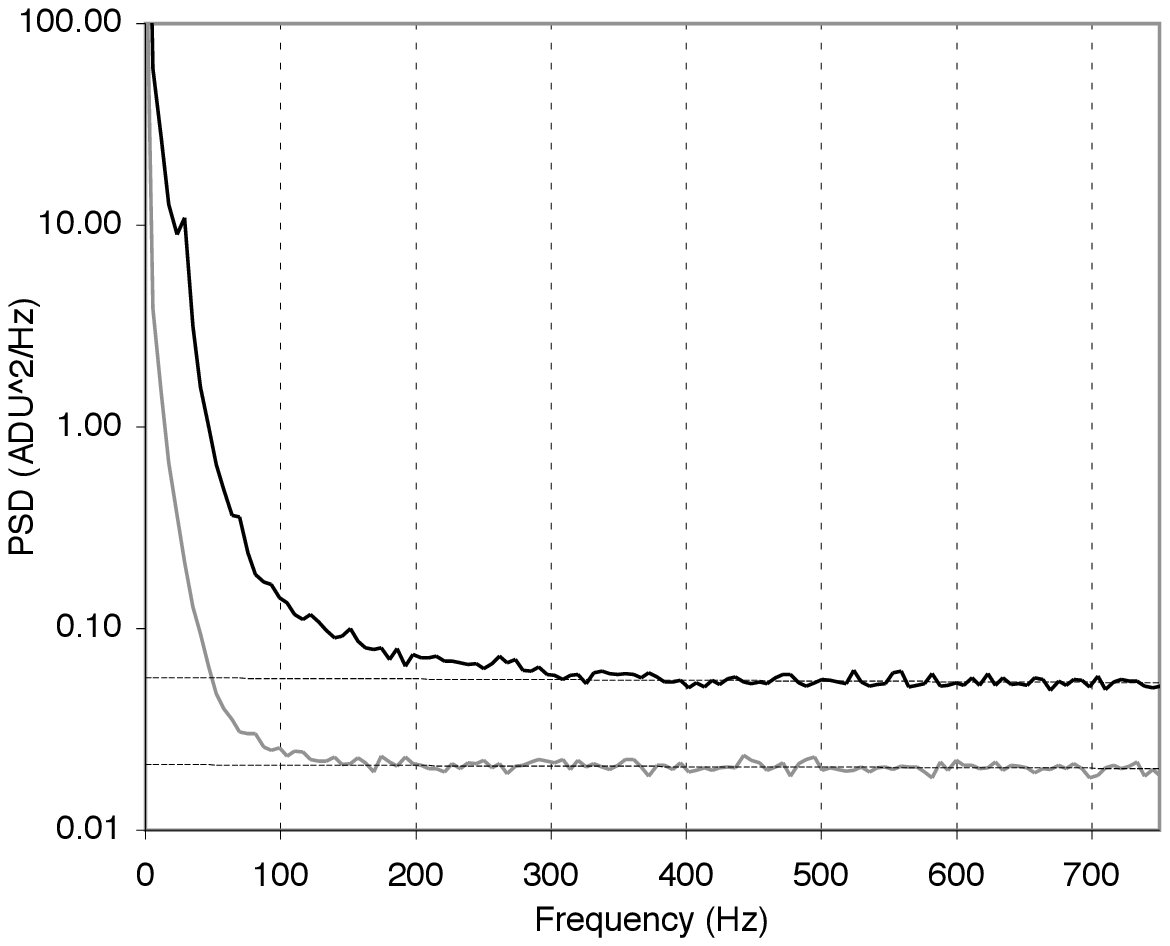}
\caption{Average power spectral density of the $P_A$ and $P_B$ signals as measured
on $\theta$\,Cen. From top to botton: $\widetilde{P_A}^2$, noise model $\widetilde{N_A}^2$
(upper dotted line), $\widetilde{P_B}^2$, noise model $\widetilde{N_B}^2$ (lower dotted line).}
\label{photometry}
\end{figure}

\begin{figure}[t]
\centering
\includegraphics[bb=0 0 360 288, width=8.5cm]{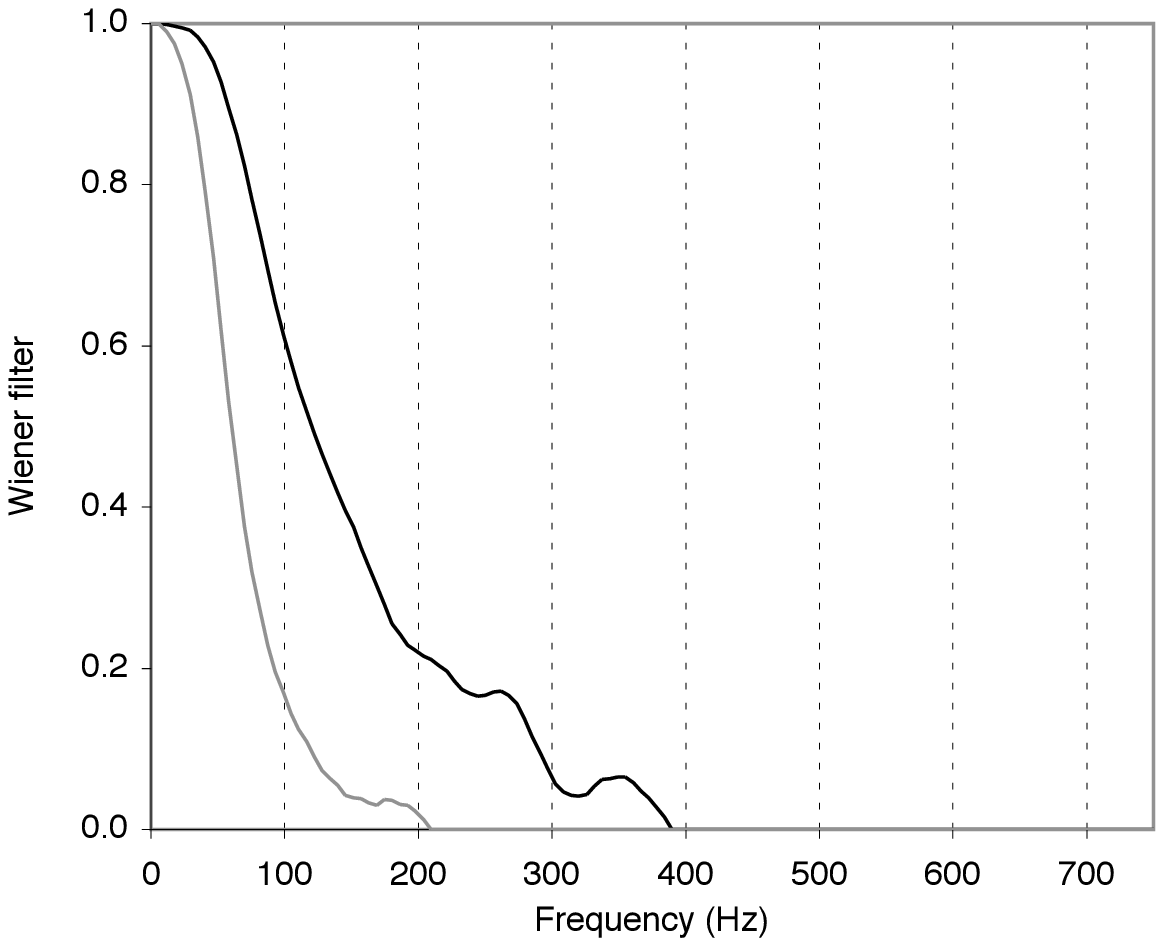}
\caption{Wiener filters computed from $\theta$\,Cen photometric signals (Fig.\,\ref{photometry}).
$W_A$ is the upper line, and $W_B$ the lower line.}
\label{wieners}
\end{figure}

\begin{figure}[t]
\centering
\includegraphics[bb=0 0 360 288, width=8.5cm]{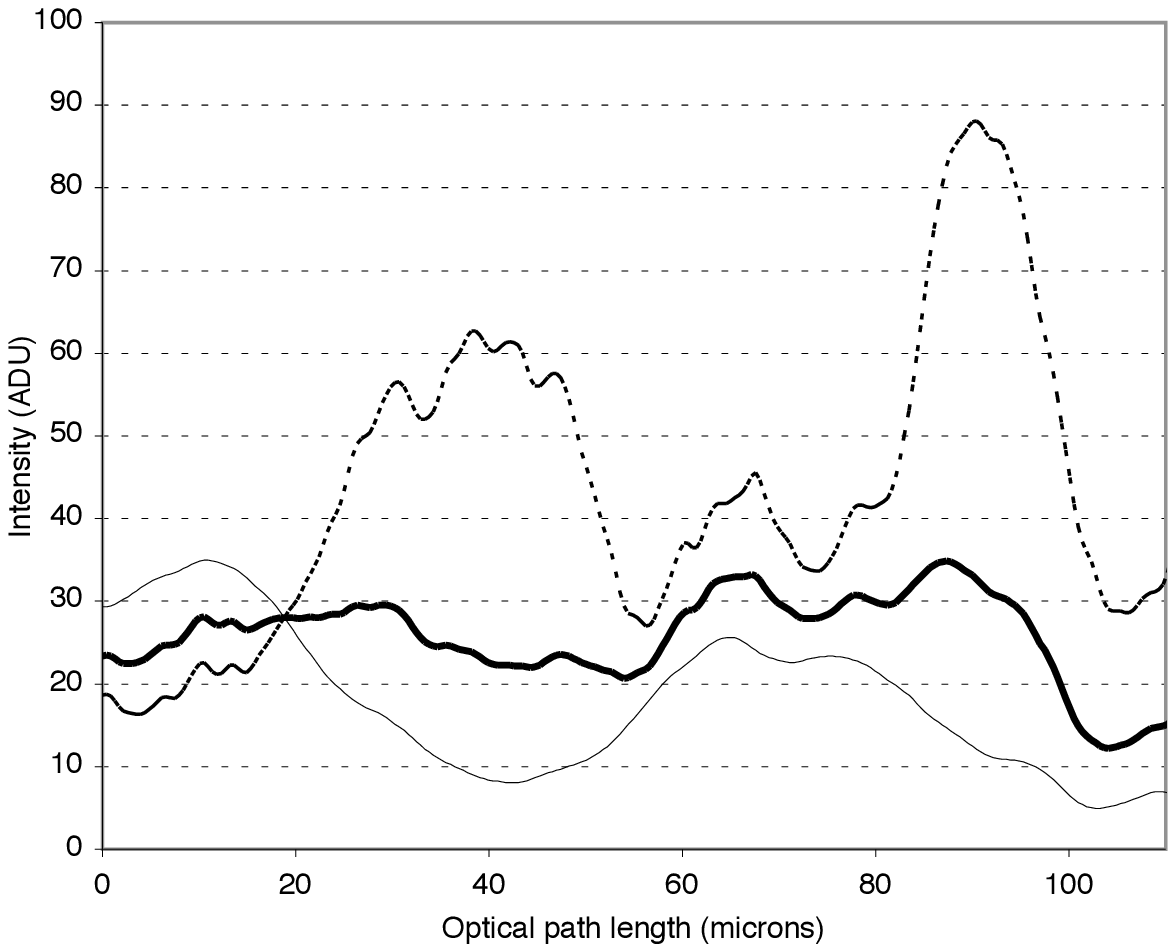}
\caption{Photometric normalization signal $P$ (thick line), with the Wiener filtered $P_A$ (dashed line)
and $P_B$ channels (thin line). $P$ is the geometric mean of the two filtered photometric channels.}
\label{P}
\end{figure}

As shown in Fig.\,\ref{photometry} ($\theta$\,Cen data), we estimate the PSD of the noise
directly from the PSD of the signal by assuming that the photon shot noise and
detector noise are constant with respect to frequency (white noise).
The contribution from the photometric fluctuations is visible at low frequencies.
Considering the high frequency part of the spectrum, we extrapolate the level of the PSD
background to the lower frequencies. Due to the high SNR of the averaged
PSD, the estimation of the background is precise enough to reconstruct $W_A$ and $W_B$.
The first frequency for which the signal becomes smaller than the noise marks
the practical limit of the Wiener filter, and the higher frequency values are set to zero.
This method is more efficient than estimating the noise PSD from the off-source batch,
as it directly takes into account the presence of photon shot noise, that also has to be removed.
The resulting Wiener filters are presented in Fig.\,\ref{wieners}.
The filtering of the photometric channels by these filters gives a clean $P$ signal,
as shown in Fig.\,\ref{P}.
\subsection{Alternative normalization methods}

If the SNR of the photometric channels $P_A$ and $P_B$ reaches too low
values over the scan length, we choose to normalize the interferograms
simply by averaging $P$ over the fringe length, instead of using the
Wiener filtered signal.
This allows us to significantly reduce the amplification of the noise due to
the normalization division. The limit between the two regimes is usually
set to 5 times the readout noise.
For interferograms that present very low photometric signal over the
fringe packet itself, we discard the scan as a significant bias
can be expected on the modulated power.
Both averaging and Wiener filtering are almost equivalent on the final calibrated
interferograms, with a slight advantage to the Wiener filtering when
the photometric fluctuations are important (as in the UT observations for example).

\subsection{Interferogram subtraction}

\begin{figure}[t]
\centering
\includegraphics[bb=0 0 360 576, width=8.5cm]{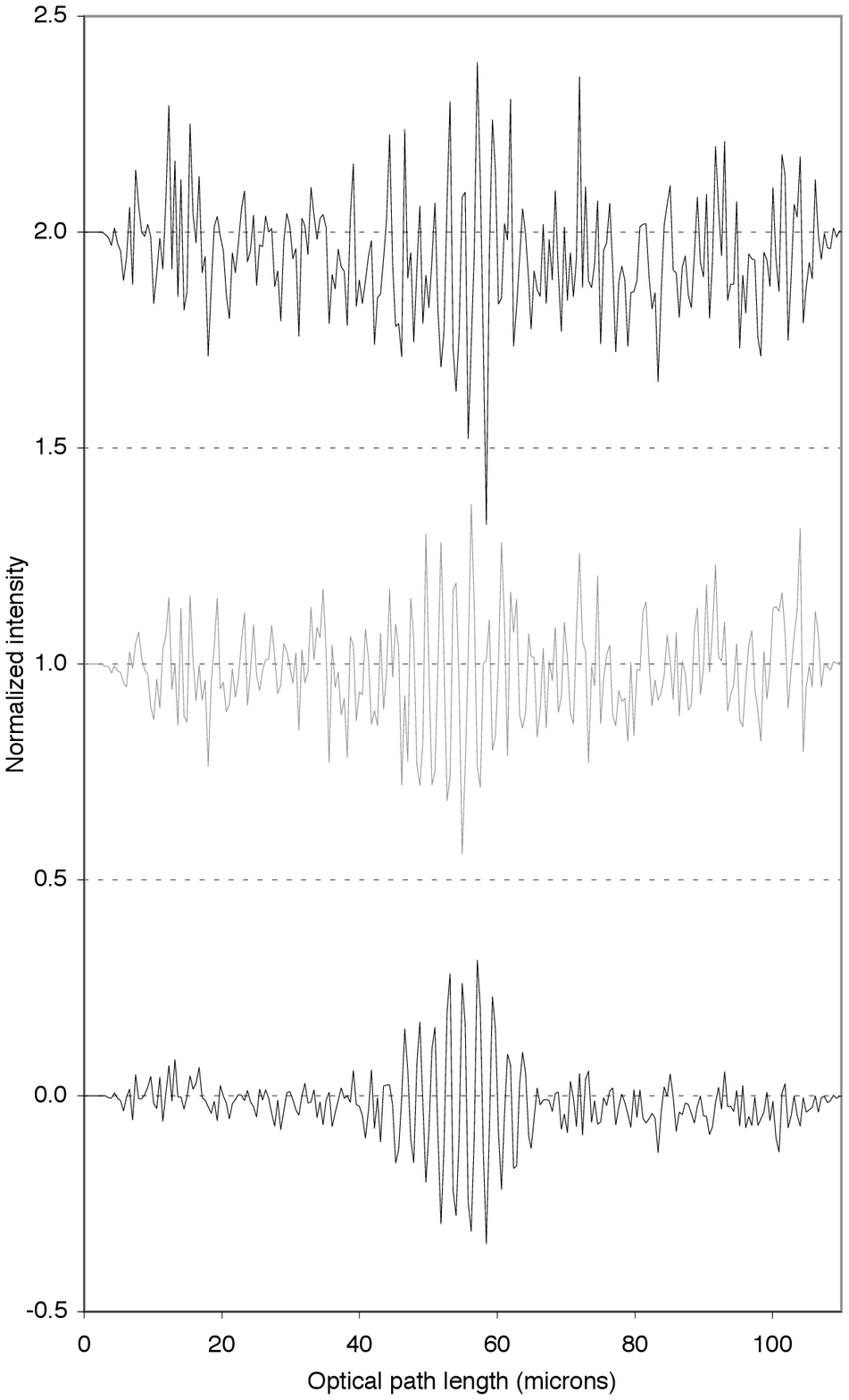}
\caption{From top to bottom: $I_1$ normalized interferogram (black), $I_2$ normalized interferogram (grey),
and the result $I$ of the subtraction of these two signals. For clarity, the $I_1$ and $I_2$ signals are shifted
vertically by +2 and +1, respectively. The correlated noise has disappeared in the combined signal and
the fringe packet appears more clearly.}
\label{I1final}
\end{figure}

After their calibration, we subtract the two interferograms $I_{1\,cal}$ and $I_{2\,cal}$,
in order to cancel the residual photometric fluctuations due to the uncertainty in
the estimation of the $\kappa$ coefficients. This subtraction has proven to significantly enhance the
immunity of the interferograms to the contamination coming from the photometric fluctuation
background.
Fig.\,\ref{I1final} shows the calibrated and normalized interferometric signal $I_{1\,cal}$ and $I_{2\,cal}$,
together with $I$ the result of the subtraction of these two signals defined as:
\begin{equation}\label{I_subtraction}
I = \frac{I_{1\,cal} - I_{2\,cal}}{2}
\end{equation}
The combined signal $I$ is used for the integration of the fringe
power to derive the visibility (Sect.\,\ref{power_integration}).
The advantage of using $I$ instead of using separately $I_{1\,cal}$ and $I_{2\,cal}$ for the integration of the
fringe power is that all the correlated noise between the two signals is eliminated by the subtraction,
while the fringes, perfectly opposed in phase, are amplified. This allows us to eliminate the residual
photometric fluctuations as well as part of the noise introduced during the photometric calibration.

\section{Interferogram selection \label{selection_interf}}

\subsection{Piston effect}

The photometric calibration of the interferograms compensates for the
incidence of wavefront corrugation across each subpupil of the
interferometer, however it does not help remove the random phase walk
(differential piston) {\em  between\/} the two subapertures.

The differential piston, considered as a time-dependent OPD
error $x(t)$, can be locally expressed by a polynomial
development around a reference time $t_{0}$ (corresponding, for
example, to the middle of the acquisition sequence):
\begin{equation}\label{piston_eq}
x(t) = x_0 + \dot{x}\ (t - t_{0}) + \ddot{x}\ (t - t_{0})^2 + ...
\end{equation}
The effect of the OPD perturbation on the interferogram, and its
consequence on the coherence factor measurement, depends on the order:
\begin{itemize}
\item {\bf Zeroth order}: the constant term $x_0$ can be seen as a
global offset of the fringe packet. It is detected and corrected by
the QL algorithm which centers the fringe packet in the
middle of the interferogram.
\item {\bf First order}:
The first order of the piston $\dot{x}$ changes the fringe velocity
and induces a simple frequency shift in the PSD. It modifies the
fringe peak position, but acts only as a homothetic compression or
expansion of the fringe packet along the OPD direction.  The first
order piston has no immediate effect on the fringe visibility.
However, if the shifting speed $\dot{x}$ of the fringe packet is too
high, it can result in an undersampling of the fringes that will
affect the visibility.
\item {\bf Second and higher orders}:
Any term of order two ({\it acceleration}) and beyond breaks the
linear relationship in the scan between time and OPD, and consequently
the Fourier relationship which is at the basis of the visibility
calculation, distorting the shape of the fringe peak.  This introduces
a non-linear, seeing induced multiplicative noise on the visibility 
measurements,
which is the dominant noise source for strong signals (bright,
unresolved objects).
\end{itemize}
Detailed studies of the properties of atmospheric piston can be found 
for example in Linfield et al.~(\cite{linfield01}) and Di Folco et al.~(\cite{difolco02}).
When a dedicated fringe tracking instrument becomes available on the VLTI
(e.g. FINITO, Gai et al.~\cite{gai03}), most of the piston will be
actively removed by a servo loop.  It remains to be checked how the
residuals will still limit the final visibility precision.

\subsection{Quality control of the interferograms \label{quality_interf}}

The goal of the interferometric data processing is to extract the squared visibility $V^2$
of the fringes. The intermediate step to this end is to measure the squared coherence
factor $\mu^2$ of the stellar light. This instrument dependent quantity characterizes
the fraction of coherent light present in the total flux of the target. It is calibrated using
observations of a known star, as described in Sect.~\ref{visibility_calibration}.
To avoid any bias on $\mu^2$, we have to reject the interferograms that
do not contain fringes (false detections), or whose fringes are severely corrupted
by the atmospheric turbulence (photometrically or by the piston effect). The selection
procedure is in practice similar to a shape recognition process.

For this purpose, we measure in the wavelet power spectral density (WPSD)
the properties of the fringe peak both in the time and frequency domains,
and we subsequently compare them with the expected properties of a
reference interferogram of visibility unity, derived from the spectral
transmission of the instrument.
In this paper, we will refer interchangeably to the ``time domain" or ``optical path
difference (OPD) domain" for the WPSD, as they are linearly related
through the scanning speed of the VINCI piezo mirror that is used to
modulate the OPD.
The fringe peak is first localized in frequency by the maximum of the
WPSD, and then the full width at half maximum is computed along the two
directions: time and frequency.
As the fringe packet has been recentered before the calibration, its position
in the time domain is zero. Three parameters are then checked for quality:
\begin{itemize}
\item peak width in the time domain (typically $\pm 50$\% around the
theoretical value is acceptable),
\item peak position in the frequency domain ($\pm 30$\%),
\item peak width in the frequency domain ($\pm 40$\%).
\end{itemize}
In principle, the variation of the fringe contrast over the
spectral band should also be taken into account to create
the theoretical reference interferogram.
But in practice, as long as the visibility of the fringes
does not cancel out for a wavelength located inside the spectral band of the
observations, the shape of the interferogram remains
very close to theoretical fringes of visibility unity.
However, when  the fringe visibility goes down to zero for a
wavelength pertaining to the observation band,
the fringe packet appears split in two parts in the time domain.
Such a deviation from the ``single packet" case can cause misidentifications,
and eventually induce a bias on the derived $\mu^2$
value, as some valid interferograms would be rejected.
In this case, one should use a dedicated broadband model to predict
the reference interferogram, taking into account the expected angular size
of the target. The selection criteria should then be adapted to match this
reference (increased peak width, presence of two maxima in the WPSD,...).
An alternative is to directly adapt the basis functions of
the wavelet transform so that they match the predicted interferograms.
We have not implemented these methods in our current
processing algorithm, whose validity is thus limited to the cases when the
visibility is above zero for all wavelengths in the observation band.
If this condition is not realized, then the interferogram selection should be
disabled, or at least the selection criteria should be made significantly less
stringent, in particular regarding the fringe peak width.
\begin{figure}[t]
\centering
\includegraphics[bb=40 70 420 270, width=8.5cm]{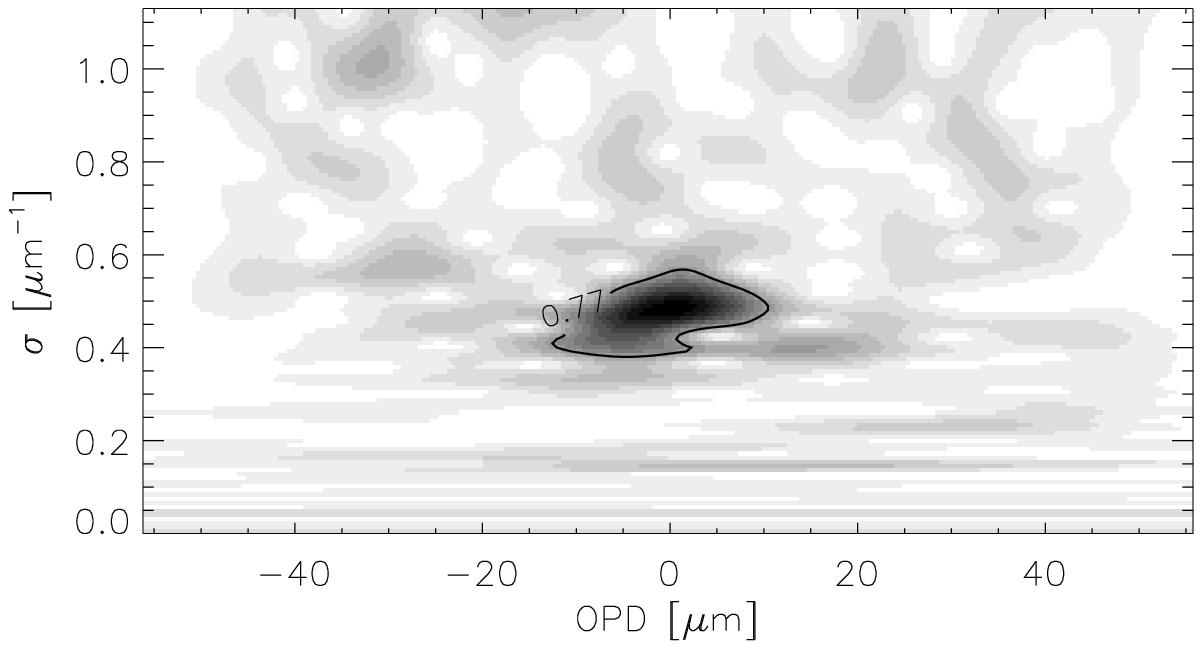}
\includegraphics[bb=40 70 420 270, width=8.5cm]{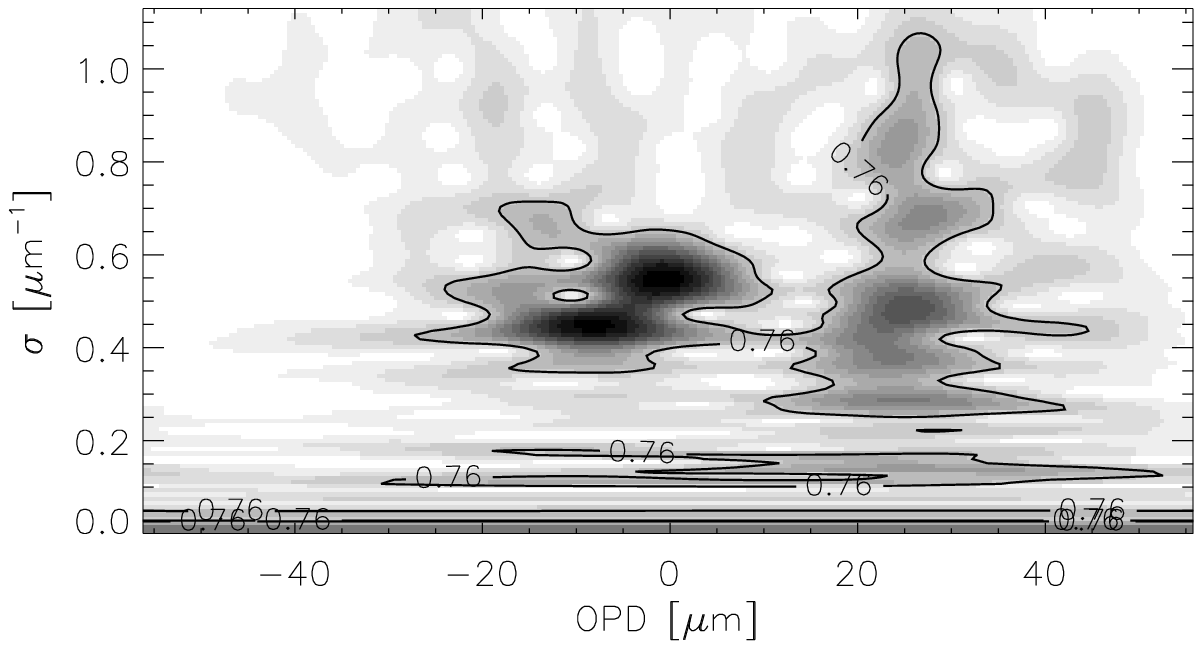}
\caption{Wavelet power spectral densities of processed interferograms
($\sigma$ represents the wave number, i.e. the inverse of the wavelength).
The upper figure shows the WPSD of a good quality interferogram:
the energy is well confied in the fringe power peak.
The bottom WPSD is affected by strong atmospheric piston:
a significant part of the fringe power is spread outside of the theoretical
fringe peak, both in time and frequency domains. The isocontours delimit
77\,\% and 76\,\% of the total modulated power, respectively.}
\label{wpsd}
\end{figure}

Fig\,\ref{wpsd} shows two examples of interferogram WPSD, one of them being
affected by atmospheric piston. The difference in terms of fringe peak shape
is clearly noticeable, and leads to the rejection of the corrupted interferogram
(bottom figure).
This selection process has shown a very low false detection rate, and rejects efficiently
the interferograms that are affected by a strong piston effect.
However, limited piston of order two (and above) is not identified efficiently.
The problem here is that the relevant properties for the estimation of the second order piston
are currently difficult to measure with a sufficient SNR from the data, as they are
masked by the order 1 piston. We expect that the introduction of the FINITO fringe
tracker in 2004 will allow us to derive an efficient metric to reject the interferograms
affected by a high order piston effect.
After the fringe power integration (described in Sect.\,\ref{power_integration}),
we filter out the scans which $\mu^2$ deviates by more than 3\,$\sigma$
from the median of the full batch of interferograms (usually 500 scans).
This step prevents the presence of very strong outliers,
which can appear due to the division introduced by the normalization of the
interferograms (introduction of Cauchy statistics).

\subsection{Immunity to selection biases \label{selection_biases}}
An essential aspect of the parameters used for the quality control
of the fringe peak properties is that they are largely independent of the
visibility of the fringes, and therefore do not create selection biases.
In particular, the integral of the fringe peak (directly linked to the visibility)
or its height are never considered in the selection. 
The parameters chosen in Sect.~\ref{quality_interf} clearly depend
on the photometric SNR, but are independent of the visibility of the
fringes, thanks to the calibration procedure described in
Sect.~\ref{photometric_calibration}.

\begin{figure}[t]
\centering
\includegraphics[bb=0 0 360 288, width=8.5cm]{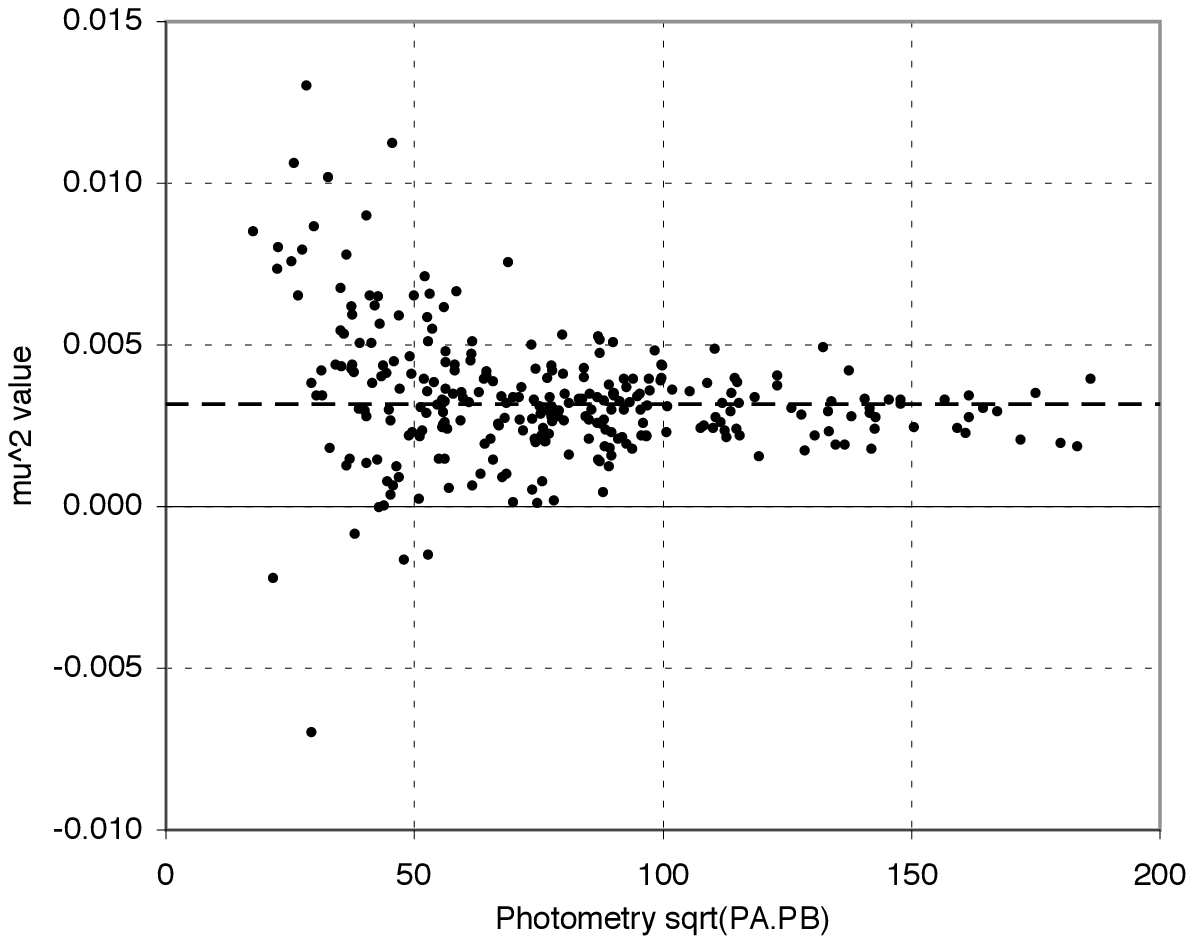}
\includegraphics[bb=0 0 360 288, width=8.5cm]{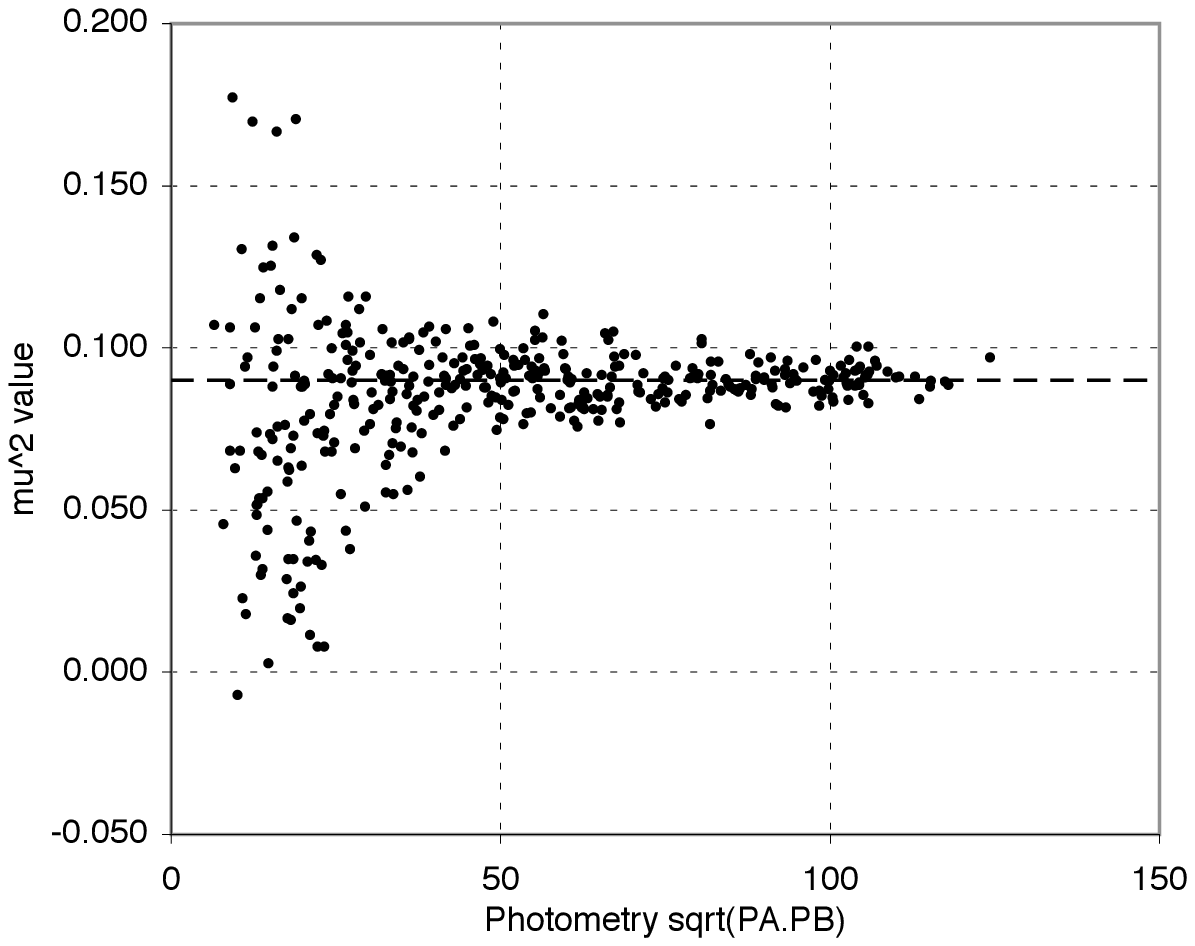}
\caption{Squared coherence factor $\mu^2$ as a function of
the photometric signal on the fringe packet, for the sample batches
obtained on $\alpha$\,Cen\,A (top) and $\theta$\,Cen (bottom).
The dashed lines show the mean value of the series derived using
the bootstrap technique.}
\label{alfcen_mu2_photom}
\end{figure}

The upper part of Fig.~\ref{alfcen_mu2_photom}
demonstrates this independence in the difficult case of the
batch of interferograms obtained on $\alpha$\,Cen\,A. Despite the
very low visibility of the fringes, no systematic deviation is visible
for low photometric SNR values, as the dispersion is symmetric
around the mean value. The same plot for $\theta$\,Cen
(Fig.~\ref{alfcen_mu2_photom} bottom) does not show
any deviation either. A further discussion of the properties of
the histogram of these measurements can be found in
Sect.~\ref{application_stars}.
This means that the quality control described in this
paragraph is not linked to the observable, and thus does not introduce
a selection bias. Its effect in the case of the $\theta$\,Cen and
$\alpha$\,Cen\,A observations is discussed in Sect.~\ref{application_stars}.

A critical case is when the visibility is extremely low.
In this situation, the fringe peak will tend to blend in with the noise, which
tends to make it appear broader and slightly displaced. Therefore,
low-visibility data are more likely to be rejected than high-visibility data.
This can introduce a bias towards higher $\mu^2$ for low-visibility
observations: a scan  with a +1\,$\sigma$ deviation is accepted, but a
scan with a -1\,$\sigma$ deviation is  more likely to be rejected as it fails
the selection criteria.
However, in this situation, the risk is high to fail to reject the
spurious spikes that are created in the calibrated interferograms
due to the division by the $P$ signal (Sect.~\ref{photometric_calibration}).
Without the selection procedure, the modulated power of these calibration artefacts
will be integrated in the final $\mu^2$ value. As this power is essentially random,
but always positive, these misidentifications would then result in a strong
positive bias on the final $\mu^2$ value. For this reason, and in spite of the potential
rejection of a small part of the valid interferograms, the application of the selection
procedure results in a more reliable estimate of $\mu^2$, even for the very low
visibility fringes. In any case, the careful examination of the statistical properties of the
$mu^2$ histogram (see Sect.~\ref{quality_control}), and in particular of its skewness,
allows us to detect a possible selection bias.
%
\section{Visibility computation}\label{power_integration}
\subsection{Wavelet power spectral density \label{wpsd_section}}
Once the interferogram has been calibrated and normalized, the squared coherence
factor $\mu^2$ is measurable as the average modulated energy of the interference
fringes over the batch.
It is computed by integrating the power peak of the interferograms in the average
WPSD (see also Appendix~\ref{morlet_appendix} and Sect.\,\ref{final_spectrum}).
The WPSD is a two dimensional matrix, examples of which are shown in Fig\,\ref{wpsd}.
For all wavelet transforms, we use the Morlet wavelet, which is defined as a plane wave multiplied
by a Gaussian envelope. It closely matches the shape of the interferometric fringe packet. 
When computing a classical PSD, the interferogram is projected on a base of sine and cosine functions, which
are not localized temporally. This means that the information of the position of the fringe packet is not used,
and that the noise of the complete interferogram contributes to the measured power.
On the other hand, the wavelet transformation projects the interferogram on a base of wavelets that are
localized both in time and frequency, making full use of the localized nature of the modulated energy. 
As discussed in Appendix~\ref{morlet_appendix}, the modulated energy of the signal is conserved
by the wavelet transform in the same way as through the classical Fourier transform.

\subsection{Estimation of the squared coherence factor $\mu^2$}

\subsubsection{Fringe power integration\label{final_spectrum}}

\begin{figure}[t]
\centering
\includegraphics[bb=0 0 360 288, width=8.5cm]{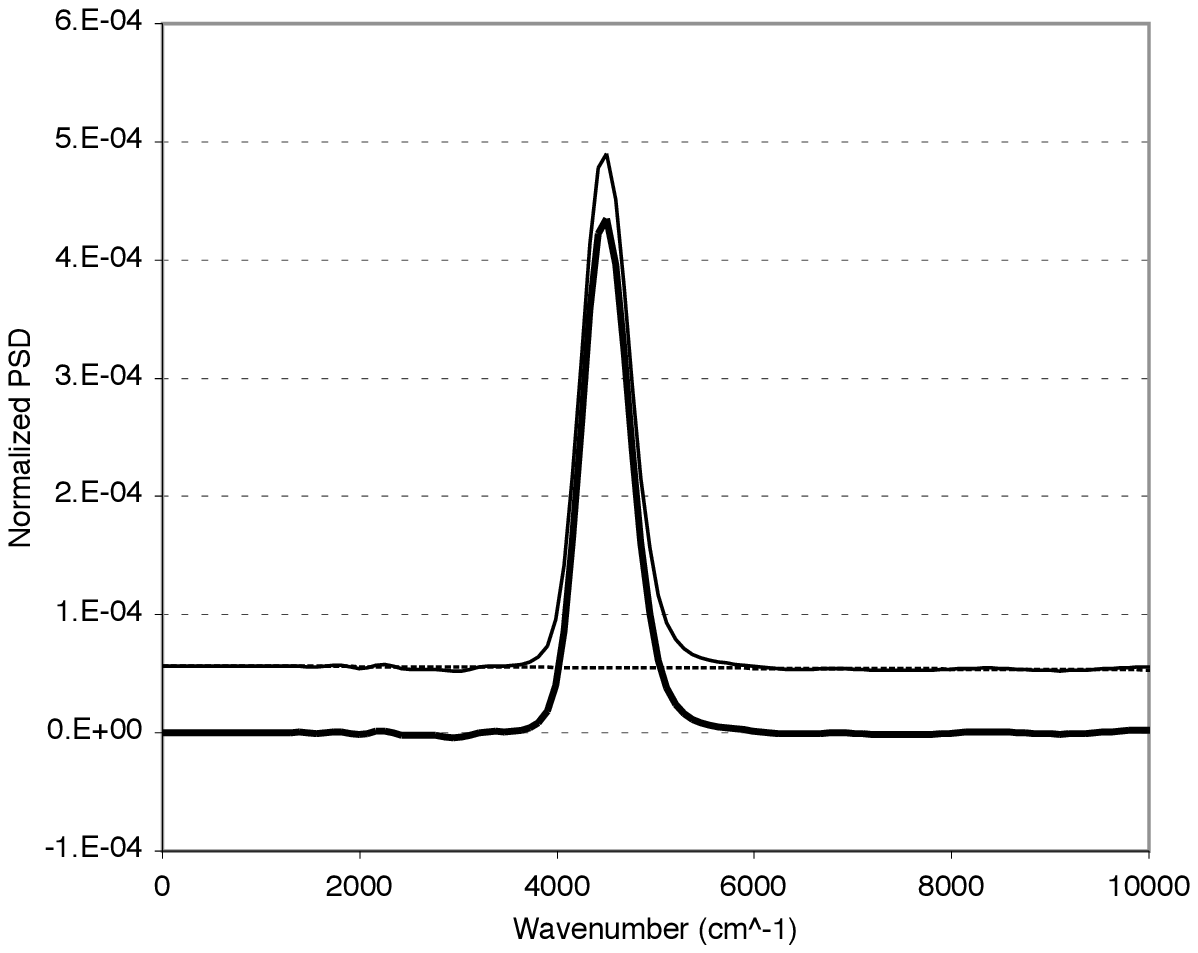}
\caption{Average WPSD of the $\alpha$\,Cen\,A observation.
The original two-dimensional matrix has been integrated over the
fringe packet length in the time domain. The resulting projection
on the wave number axis allows one to visualize clearly the noise contribution.
The subtracted noise model is shown by the dashed line.
The final WPSD (thick line) shows no bias in spite of the
large brightness of the star and very low
squared coherence factor of the fringes ($\mu^2 \approx 0.3$\,\%).}
\label{alf_cen_wl_psd}
\end{figure}

The average power spectral density of the $\alpha$\,Cen\,A sample
batch, computed using the wavelet transform, is shown in Fig.\,\ref{alf_cen_wl_psd}.
To obtain this 1D spectrum from the original 2D WPSD matrices,
we first project the WPSD matrix of each interferogram on the frequency axis, by integrating it over
the fringe packet length (time axis). From this we obtain a series of one-dimensional vector
PSDs, similar to the Fourier PSD but with a reduced noise.
Before the averaging, we recenter each fringe peak using the frequency position information
derived from the selection of the interferograms (Sect.\,\ref{selection_interf}). This step allows us
to confine  more tightly the energy of the peak, which is displaced by the first order piston effect.
This reduces the influence of piston on the final $\mu^2$ value.
The co-added 1D spectrum is the signal used for the final power
integration to estimate the $\mu^2$ of the star.

The integration of the fringe power is typically done over 100 pixels in the time
domain (20 fringes), and from 2\,000 to 8\,000\,cm$^{-1}$ in the frequency domain
(see also Sect.~\ref{fringe_pow_int}).

\subsubsection{Removal of the WPSD background\label{background_removal}}

The power in the fringe peak is contaminated by three additive components:
\begin{itemize}
\item the photon shot noise,
\item the detector noise,
\item the residual photometric fluctuations.
\end{itemize}
To estimate the modulated power of the fringes, it is essential
to precisely remove these contributions from the PSD of the interferograms.

Perrin\,(\cite{perrin03a}) has developed an analytical treatment
of the photon shot noise based on its particular properties (Poisson statistics).
The photon shot noise is perfectly white, as it is created from a purely random process.
However, due to the calibration and normalization process of the interferograms,
its translation onto the final interferometric signal $I$ could in theory deviate from
this property and show a dependence with frequency.
Such an effect has not been observed in practice on the VINCI data, 
and the uniform subtraction of the photon noise background from the PSD
of the $I$ signal has proven to be very efficient. A good example of the
"whiteness" of the photon shot noise of the processed fringes can be found in
Wittkowski et al.~(\cite{wittkowski04}), where a very bright star ($m_K = -0.6$)
was observed with the two 8\,m telescopes UT1 and UT3 ($B = 102.5$\,m).
In spite of the extremely large flux on the VINCI detector ($100$\,m$^2$ collecting optics)
and the very low visibility of the fringes ($V^2 \simeq 10^{-2}$), the resulting PSD
background is white, therefore validating our photon shot noise removal
method under the most demanding conditions.

In order to fully justify our background removal procedure,
we still have to verify the "whiteness" of the detector noise,
whose statistics and frequency structure depends on the type of detector
and readout electronics used.
The infrared camera of VINCI (LISA) is based on a HAWAII array, which is read using an IRACE controller
(Meyer et al.~\cite{irace}). As only four pixels of the 1024x1024 array are actually used, an engineering
grade detector was chosen for the instrument. It presents a large quantity of dead and hot pixels,
and therefore it was necessary to thoroughly check its noise characteristics. This was achieved
during extensive laboratory testing, and is also verified automatically for each observation.
It appeared that the LISA detector noise is perfectly white, without any significant electronic interference
signature.

This satisfactory behavior of the detector and photon shot noises
allows us to remove them simultaneously by subtracting to the $\mu^2$
value of each interferogram an average of its WPSD at high frequency,
measured outside of the domain of frequency of the interferometric fringes.
To correct for potential residuals pf the photometric calibration, we fit a
linear model of the residual background to the \emph{average} WPSD
of the interferograms in the batch.
In this procedure, we allow for a limited slope of the background model,
in order to correct a possible residual power from photometry.
Thanks to the averaging of a large number of scans, the noise on the average
WPSD low, and the fitting procedure is very precise. 
Most of the time, and even for the most important fluctuation cases (Unit Telescopes in
multispeckle mode), the contribution of the residual photometric noise is
totally negligible on the combined interferogram $I$ obtained from the subtraction
of the calibrated signals $I_{1\,cal}$ and $I_{2\,cal}$ (Eq.~\ref{I_subtraction}).

An illustration of the background quality is presented in Fig.~\ref{alf_cen_wl_psd}.
The WPSD background noise appears perfectly white, even at the very enlarged
scale used to visualize the very small fringe peak of $\alpha$\,Cen\,A ($\mu^2 \approx 0.3$\,\%).
In particular, no ``color'' or electronic interference (``pickup'') are present.
The model of the removed noise is also plotted on the two figures, showing
that the subtraction is very clean.

\subsection{Estimation of the statistical error}

To compute the statistical error on the $\mu^2$ estimation, we integrate separately
the fringe power in each WPSD of the batch, correct the detector and photon shot
noise biases individually, and use a weighted bootstrapping technique on
this set of measurements. Our sample is made of $N$ pairs $(\mu^2_i, w_i)$
where $\mu^2_i$ is the squared coherence factor obtained by integrating the WPSD of the scan
of rank $i$ in the series and $w_i$ is its associated weight. It is defined as the average level of
the photometric signal $P$ over the fringe packet length (20 fringes in the $K$ band)
multiplied by the inbalance between the two photometric channels $P_A$ and $P_B$:
\begin{equation}
w_i =   \frac{\min(\overline{P_{A,i}},
\overline{P_{B,i}})}{\max(\overline{P_{A,i}}, \overline{P_{B,i}})}
\ \overline{\left(\sqrt{P_{A,i} P_{B,i}}\right)}_{\rm Fringes}
\end{equation}
It characterizes well the clarity of the total photometric signal
that contributes to the formation of the fringes. The final dispersion
of the $\mu^2$ values is reduced by this weighting.
The detailed description of the weighted bootstrapping method used for the
computation of the error bars is given in Appendix~\ref{bootstrap_appendix}.

The bootstrapping technique has the important advantage of not making any
assumption on the type of statistical distribution that the data points follow. In particular,
it is more reliable than the classical approach that assumes a Gaussian distribution of the
measurements. Skewness and other deviations from a Gaussian
distribution are automatically included in the error bars, which can be asymmetric.

The statistical dispersion of VINCI measurements shows two regimes: for bright stars
the precision is limited by the piston and photon shot noise, while for
the fainter objects, the main contributor to the dispersion is the detector noise of the camera,
and the precision degrades rapidly. A discussion of the different types of noise
intervening in the visibility measurements can
be found for instance in Colavita~(\cite{colavita99}) and Perrin\,(\cite{perrin03a}).
The $\mu^2$ measurements discussed in this paper
have a relative statistical precision of $\pm 3.00$\%
for $\alpha$\,Cen\,A, and $\pm 0.53$\% for $\theta$\,Cen.
The lowest relative statistical dispersion $\sigma(\mu)/\mu$ reached up to now
on the coherence factor with VINCI is in the 2\% range.
Under good conditions, this translates into a bootstrapped statistical
error of less than $\pm 0.1\%$ on $\mu$ for 5 minutes of observations.

\section{Post-processing quality control \label{quality_control}}

After a batch of interferograms is processed, several quality controls are performed
in order to detect any problem in the resulting visibility values and statistical error bars. This
step is essential to ascertain the quality of the interferometric data, as it can vary depending
on the atmospheric conditions (e.g. seeing, coherence time) and on the general behavior of the
instrument (e.g. injection of the stellar light in the optical fibers, beam combiner properties,
polarization mismatch of the two beams).

\subsection{Power peak integration boundaries}\label{fringe_pow_int}

A potentially damaging effect of the atmospheric piston on the visibility of the fringes
is that it tends to move the position of the fringe
peak, and to spread it over a wider frequency range. If the frequency boundaries for
the integration of the fringe peak are set too tight, the result could be that part of the
modulated power is not taken into account, creating a bias.
These boundaries are automatically changed as a function of the ground baseline
length to account for the increased piston strength on longer baselines.
They are not modified as a function of the projected baseline, and are thus identical
for scientific targets and calibrators.

To check for the presence of such an effect, we measure the fringe peak shape
in the WPSD. More precisely, we estimate its central wave number,
full width at half maximum, as well as the limit wave numbers for which the background
level is reached.
Using these extended limits, we integrate the fringe power and compare this value
to the one obtained with the user-specified wave number limits. If a discrepancy is found at a
significant level, the batch is considered dubious and can be rejected after further
examination.
\subsection{Histogram properties}

As the noise sources acting on the $\mu^2$ values have normal statistics, it is expected
that the distribution of the $\mu^2$ values over the batch is also normal.
Although the bootstrapping procedure used to compute the $\mu^2$ error bars is
not sensitive to the type of distribution, a large skewness or kurtosis would betray a
problem in the calibration of the interferograms that could eventually bias the final
$\mu^2$ value. The relevant parameters for this verification are the skewness coefficient
$s$ (third moment of the distribution) defined as:
\begin{equation}
s = \sum_{i = 1}^{N}{\frac{(\mu^2_i-\overline{\mu^2})^3}{(N-1) \sigma^3}}\\
\end{equation}
and the kurtosis coefficient $k$ (fourth moment):
\begin{equation}
k = \sum_{i = 1}^{N}{\frac{(\mu^2_i-\overline{\mu^2})^4}{(N-1) \sigma^4}} - 3
\end{equation}
where $\mu^2_i$ are the squared coherence factor values,
$\sigma$ the standard deviation, 
$\overline{\mu^2}$ the unweighted average,
and $N$ the number of scans.
The skewness characterizes the presence of a "tail'' on the histogram.
A large value of $s$ is therefore a symptom of a potential bias problem in the
distribution, as a significant number of values are either too large or too small
compared to the average value of the sample.
A positive kurtosis coefficient means a distribution more peaked than
the normal one. However, it should be stressed that the kurtosis is not a very
robust parameter to assess if the sample is drawn from a normal distribution.
It requires a large number of sample values to be relevant,
and it is very sensitive to the presence of outliers.
Therefore, it should only be used in conjunction with other statistical
tests. A range of $\pm 0.5$ can be considered acceptable for $k$.
When random samples are drawn from a normal population,
the resulting skewness coefficients will fall into the range $\pm 0.18$,
with a probability of 90\%. We therefore choose this value as an acceptable range.

\subsection{Application to $\theta$\,Cen and $\alpha$\,Cen~A \label{application_stars}}

Table~\ref{rejection_reasons} gives the reasons for the rejection
of the interferograms of the $\theta$\,Cen and $\alpha$\,Cen~A batches.
In the case of $\alpha$\,Cen\,A, a larger number of interferograms are
rejected due to the very low visibility of the fringes.
\begin{table}
\caption[]{Reasons for the rejection of $\theta$\,Cen and $\alpha$\,Cen~A
interferograms during the processing. The lower part of this table
corresponds to the selection criteria related to the atmospheric piston effect.}
\label{rejection_reasons}	
\begin{tabular}{lcc}
\hline
Reason & $\theta$\,Cen & $\alpha$\,Cen~A \\
\hline
Photometry too low & 77 & 24 \\
Large OPD jump & 13 & 47 \\
Fringes at edge & 6 & 27 \\
Fringe packet width & 1 & 47 \\
Fringe peak position & 3 & 40 \\
Fringe peak width & 5 & 33 \\
Outliers (3\,$\sigma$) & 9 & 5 \\
\hline
\noalign{\smallskip}
Total number of rejected scans & 114/500 & 223/500 \\
\noalign{\smallskip}
\hline
\end{tabular}
\end{table}

The measured statistical properties 
of the processed interferograms of $\theta$\,Cen and $\alpha$\,Cen\,A
are given in Table~\ref{stat_properties}.
The values in brackets were obtained by disabling the piston selection
of the interferograms (based on the fringe packet width, and on the
position and width of the fringe peak in the power spectrum of the
interferogram). The comparison of the selected vs. non-selected versions
of the data processing shows that the piston selection has a positive effect
on the dispersion of the measurements. For $\theta$\,Cen, the
difference is minimal between the two kinds of processing. In particular,
the total number of processed interferograms is almost identical for the two
cases. However, for $\alpha$\,Cen\,A, the difference is clearly noticeable,
as the final error bars are 60\% larger when the selection is disabled,
in spite of a total number of processed scans approximately 40\% larger.
The skewness of the histogram is also much larger in this case (by a factor of 20).
This clearly shows the advantage of the fringe selection procedure, in
particular for the rejection of the calibration artefacts (false detections)
in the very low visibility case (see also Sect.~\ref{selection_biases}).

\begin{table}
\caption[]{Statistical properties of the $\theta$\,Cen and $\alpha$\,Cen\,A
sample batches. The values obtained when disabling
the selection of the interferograms based on their piston properties
are given in brackets for comparison.}
\label{stat_properties}\begin{tabular}{lcc}
\hline 
& $\theta$\,Cen & $\alpha$\,Cen\,A\\
\hline
Reduced scans & 386 (395) & 277 (391) \\
Average $\mu^2$ (\%) & 8.995 (8.999) & 0.3180 (0.2949)\\
Stat. error ($1\sigma$) & 0.048 (0.050)& 0.0095 (0.0142) \\
Rel. error $\sigma/\mu^2$& 0.53\% (0.56\%) & 3.00\% (4.81\%) \\
Skewness $s$ & $+0.023$ ($+0.013$) & $-0.007$ ($+0.160$) \\
Kurtosis $k$ & $+0.164$ ($+0.062$) & $+0.044$ ($-0.042$) \\
\hline
\end{tabular}
\end{table}

\begin{figure}[t]
\centering
\includegraphics[bb=0 0 360 144, width=8.5cm]{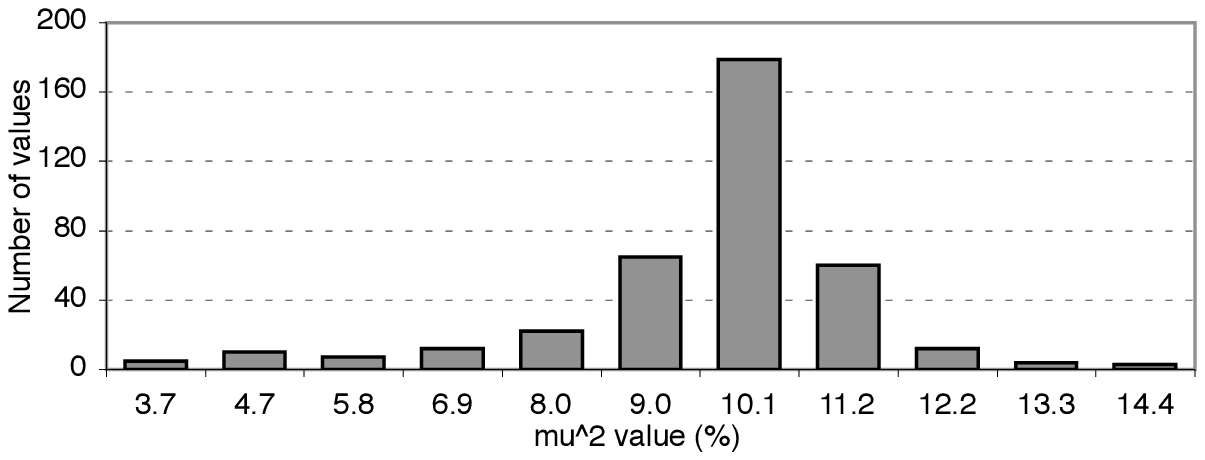}
\caption{Histogram of the $\mu^2$ values obtained on $\theta$\,Cen. No
significant skewness is present.}
\label{theta_cen_histogram}
\end{figure}

\begin{figure}[t]
\centering
\includegraphics[bb=0 0 360 144, width=8.5cm]{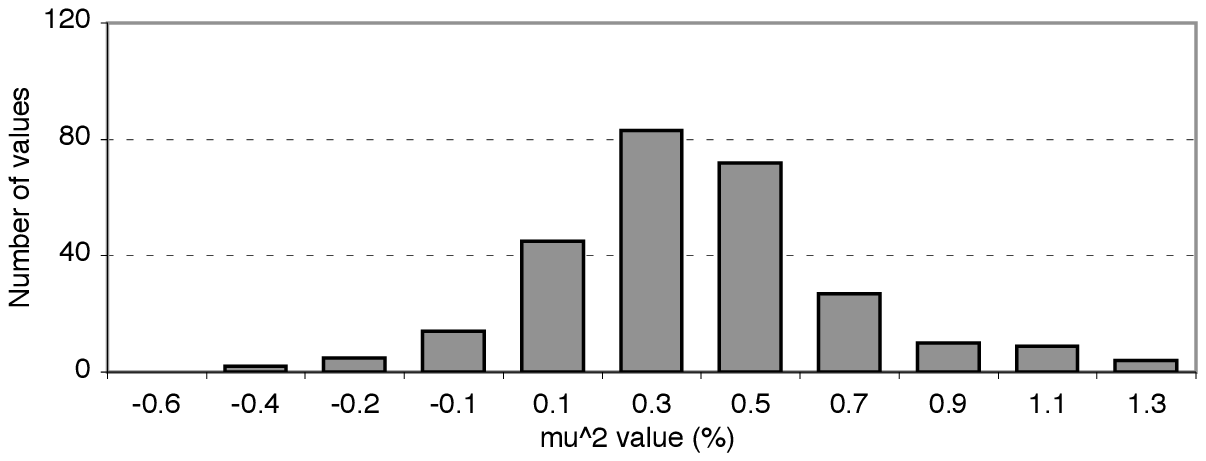}
\caption{Histogram of the $\mu^2$ values obtained on $\alpha$\,Cen\,A.
A moderate skewness is present towards higher $\mu^2$ values.}
\label{alf_cen_histogram}
\end{figure}

In the case of $\theta$\,Cen (Fig.\,\ref{theta_cen_histogram}) and
$\alpha$\,Cen\,A (Fig.\,\ref{alf_cen_histogram}), no skewness is present.
For $\theta$\,Cen, a small positive kurtosis $k \approx 0.16$
is detected, meaning that the distribution is slightly too peaked ({\it leptokurtic},
as opposed to a {\it platykurtic} distribution that is too flat).
However, it is easily inside the acceptable range ($\pm 0.5$), and this property is
taken into account in the bootstrapped error bars.


\section{Visibility calibration \label{visibility_calibration}}
\subsection{Principle}

The data reduction software of VINCI yields accurate estimates of the squared
modulus of the coherence factor $\mu^{2}$, which is linked to the squared
object visibility $V^2$ by the relation:
\begin{equation}
V^{2} = \frac{\mu^2}{T^2}
\end{equation}
where $T^2$ is the response of the system to a point source,
also called transfer function (hereafter TF),
interferometric efficiency, or system visibility.
It is measured by bracketing the science target with observations of calibrator
stars whose $V^2$ is supposed to be known a priori.
The accuracy of our knowledge of the calibrator angular diameter, and the precision with
which we estimate its $\mu^2$ are therefore decisive for the final quality of the
scientific target observation.
Typically, the scientific targets are bracketed by calibrator observations,
so as to be able to verify the stability of the TF.
In this respect, the VLTI has often proved to be stable at a scale of a few percent over
several nights. Nevertheless, and to guarantee the quality of the VINCI data,
calibrators are regularly observed during the night, before and after each scientific target.

\subsection{Interferometric transfer function estimation}

\subsubsection{Transfer function model \label{TF_model}}

By nature, the interferometric TF is affected by a large number of parameters:
atmospheric conditions (seeing, coherence time), polarization (incidence of the
stellar beams on the siderostat mirrors, spectrum of the target, etc...).
These effects combine to make $T^2$ a stochastic
variable, that can evolve over a wide range of timescales.
In order to estimate its value and uncertainty on a particular date at which
it was not directly measured (e.g. during the observation of a scientific target),
it is necessary to use a model of its evolution. Such a model relies necessarily
on an hypothesis, for instance that the value of $T^2$ is constant between
two (or more) calibrator observations, that it varies linearily, quadratically,
or any higher order model. Let us now evaluate the most suitable type of
TF model for the observations with VINCI.

\begin{figure}[t]
\centering
\includegraphics[bb=0 0 360 288, width=8.5cm]{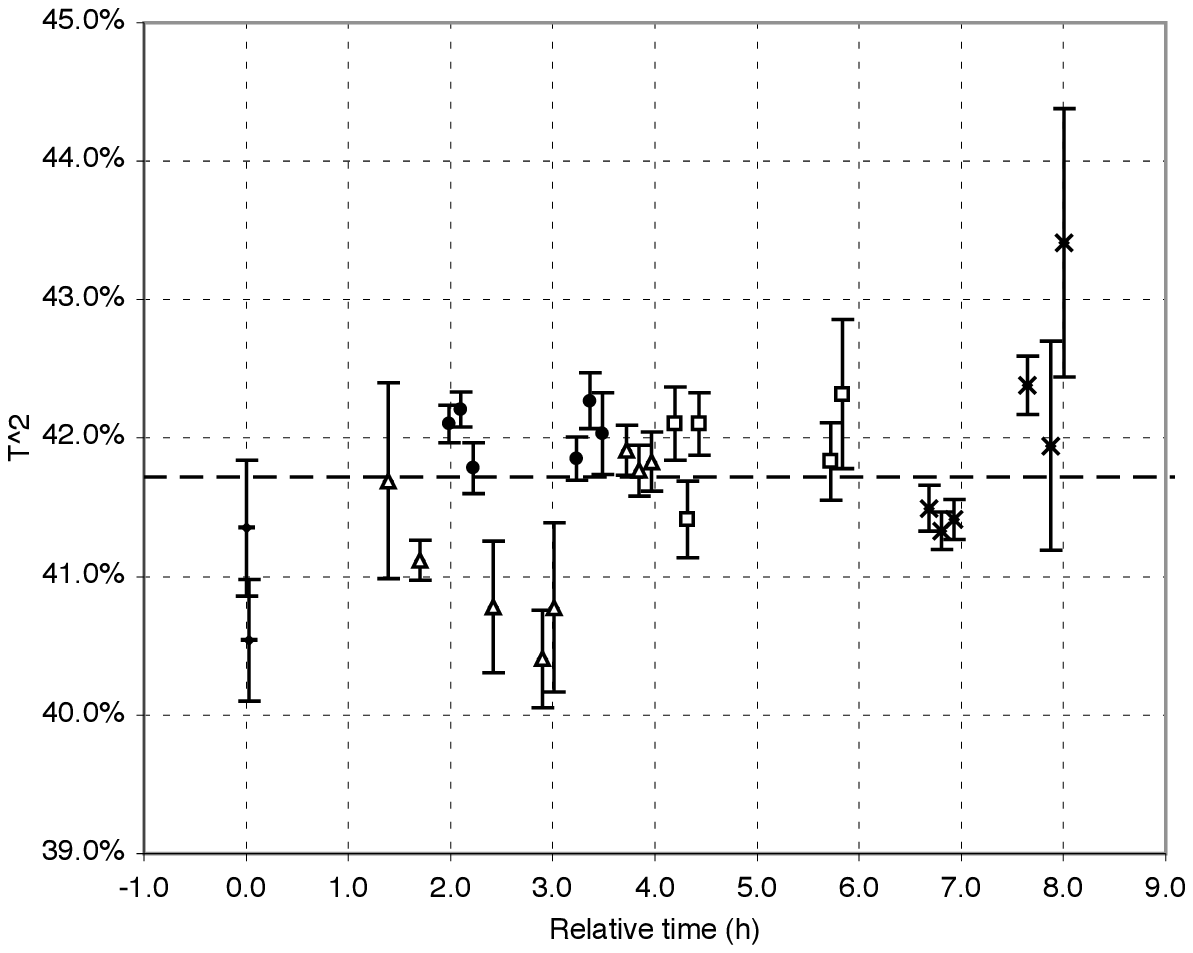}
\caption{Evolution of the transfer function $T^2$ during one night (2002-05-29)
on the E0-G0 baseline of the VLTI (16\,m in ground length).
Each symbol corresponds to a different star.}
\label{calibrators_night}
\end{figure}

As a practical example, Fig.~\ref{calibrators_night} shows the evolution of
$T^2$ over one night of observations, with a typical
sampling rate of one measurement every 15 minutes.
This series of 27 observations was obtained during the night of 29 May 2003
on the E0-G0 baseline (16\,m ground length).
A number of different stars with known angular diameters were observed,
covering spectral types in the G-K range.
During these observations, spread over 8 hours, the seeing evolved from
1.0 to 2.0 arcsec, the altitude of the observed objects was
distributed almost uniformly between 25 and 80 degrees,
and the azimuth values covered 15 to 90 degrees (N = 0, E = 90).
Due to this broad range of conditions, this series represents
a worst case in terms of TF stability. As a reminder, under
normal conditions, a calibrator is selected as close as possible to the
scientific target, in time, position and spectral type.

Over the whole night, the overall stability is satisfactory, with a
dispersion of $\sigma_{\rm tot} = 0.64$\% around the average
value of $\overline{T^2} = 41.75$\%.
In order to estimate the external dispersion $\sigma_{\rm ext}$ of the transfer
function over the night (due to the atmosphere and instrumental drifts), we can
subtract the average of the intrinsic variances $\sigma_i^2$ of the $T_i^2$ values
$\sigma_{\rm int}^2$ from the total variance $\sigma_{\rm tot}^2$:
\begin{equation}
\sigma_{\rm ext}^2 = \sigma_{\rm tot}^2 - \sigma_{\rm int}^2.
\end{equation}
The average precision of each individual $T^2$
measurement in our sample night is $\sigma_{\rm int} = 0.21\%$.
This gives an external dispersion of $\sigma_{\rm ext} = 0.60\%$.
In this particular case, the external dispersion is thus dominant over the
internal measurement errors, by a factor of almost three.

From this example, we can conclude that the rate of one measurement
every 15 minutes is insufficient to sample the fluctuations of the TF.
Due to this, we do not gain in precision by interpolating the TF values
using a high order model (quadratic, splines,...).
In the current state of the VLTI (siderostat observations), the most adequate
model for the estimation of the TF is thus a constant value between the observations
of the calibrators.
The 1.8\,m Auxiliary Telescopes will soon allow us to sample
the TF with a much higher rate, of the order of 1 minute, and
higher order models of the TF variations could become necessary.
As we are dominated by the external dispersion $\sigma_{\rm ext}$,
the uncertainty on the TF has to be estimated from the dispersion of
the individual $T^2$ measurements obtained before and after the scientific
target, without averaging of their associated error bars.

\subsubsection{Refinement of the transfer function model}

Under good and stable conditions, the random dispersion of $T^2$ introduced
by the atmosphere can be very low between two consecutive observations of a
calibrator. In this case, we want to evaluate the true uncertainty on the model $T^2$
by comparing the hypothesis of stability to the calibrator observations,
and subsequently refine the hypothesis used to estimate the error bar on $T^2$.

The observational strategy chosen with VINCI is to record several
series of interferograms consecutively for each calibrator observation (typically three),
over a period of about 15 minutes.
To decide if the atmospheric and instrumental conditions are stable over
this period, we compute the following $\chi^2$ expression:

\begin{equation}
\chi^2_{\rm tot} = \sum_{i}^{} \frac{(T_i^2 - \overline{T^2})^2}{\sigma_{{\rm stat},i}^2}
\end{equation}
where $T_i^2$ are the consecutive estimates of the TF obtained on the calibrator,
$\sigma_{{\rm stat},i}^2$ the statistical error of each measurement,
and $T^2$ the weighted average of the $T_i^2$ values
(using the inverse of the statistical variance as weights).
If the resulting $\chi^2$ is small (less than 3), then the hypothesis that the TF is
constant is probably true: the $T_i^2$ values can be averaged and the global
statistical error bar reduced accordingly. If not, then this hypothesis cannot be made,
and a realistic approach is to consider as the true measurement error of the average
TF the standard deviation of the $T_i^2$ sample.

When several series of interferograms are obtained on the same calibrator and the conditions
described above are verified, the resulting estimates of the TF can be averaged in order to
reduce the attached statistical error bar. However, the systematic error
introduced by the {\it a priori} uncertainty on the angular size of one calibrator
cannot be reduced by repeatedly observing this star, but only by combining the
TF measurements obtained on independent objects.

\subsection{Application to the sample observation of $\alpha$\,Cen\,A}

In the case of the observations described in this paper, $\theta$\,Cen was
observed one hour before $\alpha$\,Cen\,A. Assuming a UD angular diameter
of $\theta_{\rm UD} = 5.305 \pm 0.020$~mas (Kervella et al.~\cite{kervella03b}),
and taking into account the spectrum of the source and the bandwidth averaging
effect (also called bandwidth smearing, see e.g. Davis et al.~\cite{davis00}),
we expect a squared visibility of $V_{\rm theo}^2 = 0.1796$ and a $1\,\sigma$
systematic uncertainty $\sigma_{\rm syst} = \pm 0.0027$ for the
65.929\,m projected baseline (weighted average over the interferogram series).
As we observed an instrumental coherence factor of $\mu^2 = 0.08995 \pm 0.00048$,
the transfer function $T^2$ is estimated to be:
\begin{equation}\label{T2_eq}
T^2 = \frac{0.08995\ [\pm 0.00048]_{\rm stat}}{0.1796\ [\pm 0.0027]_{\rm syst}}.
\end{equation}
Using the relations detailed in Appendix~\ref{appendix_ratio} to compute the error bars
of the ratio $\mu^2/T^2$, we obtain
\begin{equation}
T^2 = 0.5010\ [\pm 0.0027]_{\rm stat}\ [\pm 0.0077]_{\rm syst}.
\end{equation}
As we consider here only one calibrator observation, we cannot estimate the external dispersion of
$T^2$, and we consider only the internal statistical and systematic error bars.
As a remark, this $T^2$ value is not identical to the one computed in
Sect.~\ref{TF_model}, as it was obtained more than one month later.
The VINCI coupler is known to be sensitive to long term
temperature variations (over a timescale of weeks), an effect that can explain the
observed difference.

The squared visibility value of $\alpha$\,Cen\,A is then:
\begin{equation}
V_{\rm \alpha\,Cen}^2 = 0.00635\ [\pm 0.00019]_{\rm stat}\ [\pm 0.00010]_{\rm syst}.
\end{equation}
The uncertainty on this value is dominated by the statistical error, despite the importance
of the systematic error on the value of $T^2$. 
The average baseline of this measurement is 61.470~m, and we can now deduce
the uniform disk model angular diameter of $\alpha$\,Cen\,A, $\theta_{\rm UD} = 8.305 \pm 0.024$\,mas,
which is very close to the published value of $\theta_{\rm UD} = 8.314 \pm 0.016$~mas from
Kervella et al.~(\cite{kervella03b}). This computation takes into account the
wavelength averaging effect due to the broadband $K$ filter of VINCI
as described by Kervella et al.~(\cite{kervella03b}).

\section{Conclusion}

We have described the data reduction methods that are used on VINCI, the VLTI
Commissioning Instrument.
In particular, we detailed the photometric calibration of the interferometric signals,
followed by the normalization of the fringes, and the subtraction of the
two calibrated interferograms.
Due to the efficient spatial filtering provided by the single-mode optical fibers,
this procedure provides a clean calibration of the fringes, and allows us to derive the
squared coherence factor $\mu^2$ with high accuracy.
Combined with observations from a calibrator star, it yields the squared visibility $V^2$.
This value can be interpreted physically through the use of a dedicated model of the
observed object. 
Applying the data reduction methods described in this paper
to sample data from $\alpha$\,Cen\,A yields a realistic value of its uniform
disk angular diameter.
Our procedures can easily be adapted to other single mode interferometric instruments.
In particular, they can be generalized to spectrally dispersed fringes and to a multiple beam
recombiner using the integrated optics technology (Kern et al.~\cite{kern03}).
Such a device could allow the simultaneous recombination of the beams from
the four 8\,m Unit Telescopes and four Auxiliary Telescope of the VLTI in
a compact instrument.
%


\appendix

\section{The wavelet transform\label{morlet_appendix}}

The wavelet transform belongs to the class of time-frequency transforms which are
powerful tools to study non-stationary processes such as turbulent flows in fluid mechanics.  
Wide band coaxial interferograms recorded through a turbulent atmosphere can
be strongly distorted due to the differential piston effect and fast photometric
fluctuations. In this context, the wavelet transform is an efficient tool to study and
analyse the interferograms recorded from the ground.

The continuous wavelet transform (hereafter CWT) is defined by:
\begin{equation}
W\left(s,\tau\right)=\frac{1}{\sqrt{s}} \int_{-\infty}^{+\infty}
f\left(t\right) \psi^{*}\left(\frac{t-\tau}{s}\right) \,dt
\label{weq-0}
\end{equation}
where $f(t)$ is the signal defined as a function of time,
$\psi\left(t\right)$ the chosen wavelet function, $\psi^{*}$
its complex conjugate, $s$ the scale, and $\tau$ the translation.

\begin{figure}[t]
\centering
\includegraphics[bb=0 0 360 288, width=8.5cm]{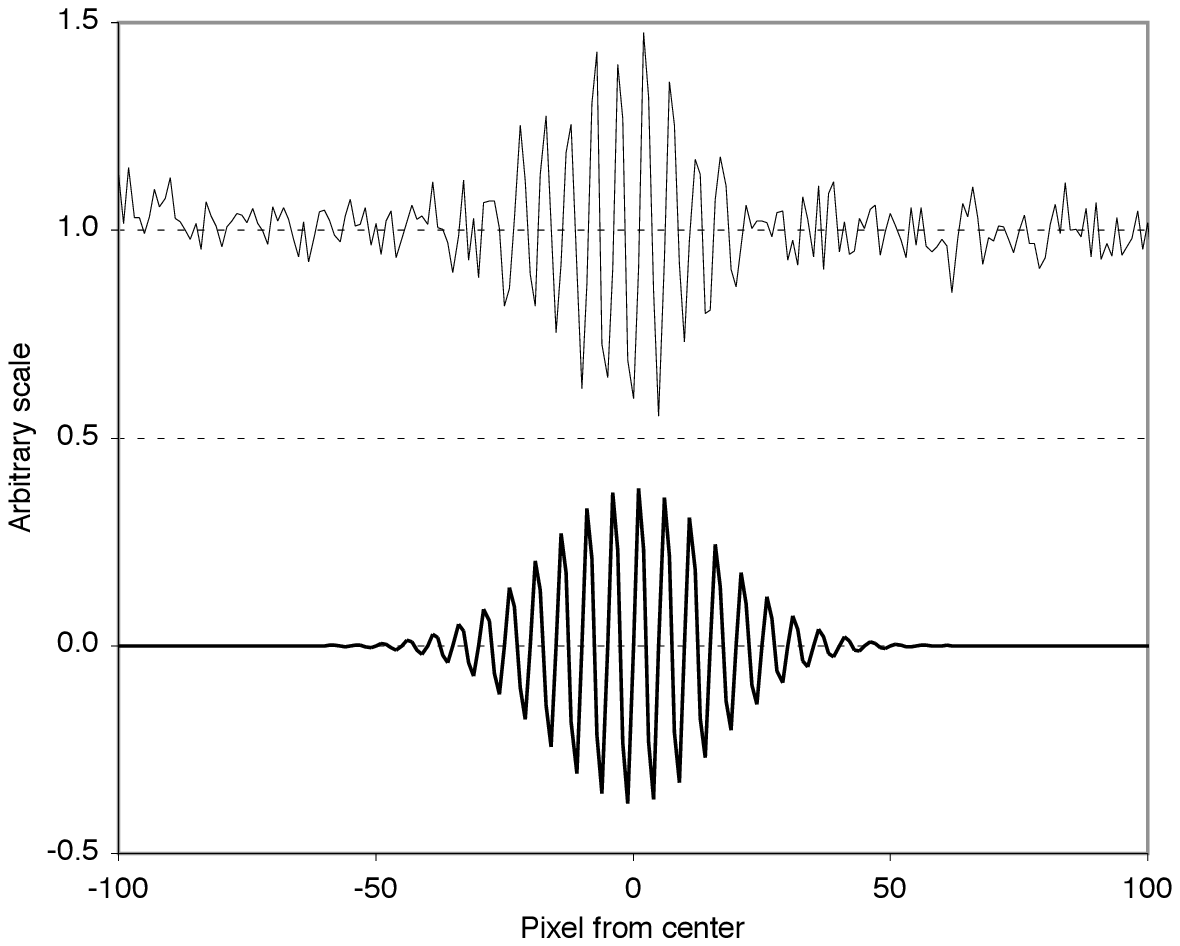}
\caption{VINCI interferometric fringes (upper curve, from a processed
interferogram of $\theta$\,Cen) and the Morlet wavelet function imaginary
part (bottom curve).}
\label{Morlet_wl}
\end{figure}

For the present application of the CWT to interferometry, we have chosen to
use the Morlet wavelet, that is defined as a Gaussian envelope multiplied
by a plane wave (Goupillaud et al.~\cite{goupillaud84}; Farge~\cite{farge92}):
\begin{equation}
\psi\left(\eta\right)= \exp{( i 2 \pi \nu_{0} \eta )}\ \exp{(-\eta^2 / 2)}
\label{weq-1}
\end{equation}
where $\eta$ is the non dimensional time parameter and $\nu_{0}$
is the non dimensional frequency.
Initially used for the analysis of seismic signals, the
Morlet wavelet is a good approximation of the fringe pattern produced by VINCI.
The data processing methods presented here make extensive
use of this particular wavelet for the recognition and localization of the fringes
(Sect.~\ref{fringe_localization}) and subsequently for the integration of the
modulated power of the interference fringes (Sect.~\ref{power_integration}).
Fig~\ref{Morlet_wl} shows the shape of the imaginary part
of the Morlet wavelet assuming typical parameters for the processing of data from
the MONA beam combiner ($K$ band).

If we now express the CWT in the Fourier domain (Eq.~\ref{weq-2}), it appears
clearly that the CWT is a filtered version of the signal for different sets of filters:
\begin{equation}
W(s,\tau)= \sqrt{s}
\int_{-\infty}^{+\infty} \widehat{f}\left(\nu\right)
\widehat{\psi}^{*}\left(s\nu\right) e^{i\,2 \pi \nu \tau}\,d\nu
\label{weq-2}
\end{equation}
Since the CWT is simply a convolution between the signal $f\left(t\right)$ and
expanded/contracted versions of the wavelet function (Eq.~\ref{weq-0}), the Morlet
wavelet is very efficient to analyse wide-band coxial interferograms. 
The CWT of an interferogram using the Morlet wavelet is a complex quantity
and its maximum energy is found for the wavelet that is most similar to the
recorded interferogram.

The CWT using the Morlet wavelet is not orthogonal but since it relies on
a set of filtered versions of the signal with strong redundancy, the original
signal can easily be reconstructed (Farge~\cite{farge92}; Perrier~\cite{perrier95}).
The energy properties of Wavelets are similar to the ones of the Fourier analysis,
with the equivalent of the Parseval theorem (Perrier~\cite{perrier95}).
We have therefore the equivalence of the two following expressions of
the energy $E$ of the signal:
\begin{equation}
E =\frac{1}{2\,C_{\psi}}  \int_{0}^{+\infty} \int_{-\infty}^{+\infty}
\left| W(s,\tau) \right|^{2}\, \frac{ds}{s^{2}} \,\, d\tau
\end{equation}
\begin{equation}
E = \int_{-\infty}^{+\infty}\left|  \widehat{f}\left(\nu\right)\right|^{2}\,d\nu
\label{weq-3}
\end{equation}
with the coefficient $C_{\psi}$ defined as:
\begin{equation}
C_{\psi}=\int_{-\infty}^{+\infty} \frac{\left|\psi\left(s\,\nu\right)\right|^{2}}{\nu}\,d\nu 
\end{equation}
As a consequence, we are able to recover the modulated energy of the
original signal (the interferometric fringes) by integrating its wavelet power
spectrum over the time and frequency regions where
the interferogram is present.

Compared to the classical Fourier analysis, such an approach allows
to minimize the biases due to both the white and colored
(frequency dependent) noises. Thanks to its localization
both in time and frequency, the Morlet wavelet is better suited to the study of
interferometric fringe packets than the classical Fourier base functions
(sine and cosine), as the noise present outside of the fringe
packet in the scan is excluded from the integrated power.
The interested reader will find a more detailed treatment of the wavelet transform
in Daubechies~(\cite{daubechies92}), Farge~(\cite{farge92}),Perrier~(\cite{perrier95})
and Mallat~(\cite{mallat99}).

\section{Computation of weighted bootstrapped error bars \label{bootstrap_appendix}}

Originally developed by Efron\,(\cite{efron79}), the bootstrap analysis,
also called sampling with replacement, consists of constructing a hypothetical, large
population derived from the original measurements and estimate the statistical properties from
this population. This technique allows us to recover the original distribution
characteristics without any assumption on the properties of the underlying population
(e.g. Gaussianity). An introduction to bootstrap analysis can be found in
Efron\,(\cite{efron93}) and Babu\,(\cite{babu96}).

Our implementation of the bootstrapping technique draws, {\emph with repetition},
a large number $M$ of  samples containing $N$ elements from the original set
of measurements $(\mu^2_i, w_i)$, also $N$ elements in length. $\mu^2_i$ designates
the squared coherence factor associated with the scan of rank $i$ in the series, and
$w_i$ is its associated weight.
The result of this drawing is an $N \times M$ matrix of $(\mu^2_{k,j}, w_{k,j})$ pairs
($1 \le k \le N$; $1 \le j \le M$).
The fact that the same element of the
original data set can be repeated several times in the drawing is essential,
as it allows us to create independent samples.
Typically, several thousand samples are obtained from the original data, which
contains a few hundred $\mu^2$ values. The weighted average values $\mu^2_k$
are computed for each of the $N$ drawings:
\begin{equation}
\mu_j^2 = \frac{1}{\left(\sum_{k=1}^{N} w_{k,j}\right)} \sum_{k=1}^{N} w_{k,j} \mu_{k,j}^2
\end{equation}
The resulting ensemble $[\mu^2_j]$ ($M$ elements) is sorted in ascending
order, and reindexed with the percentiles of the rank of each value in the set:
\begin{equation}
\mu^2_{0/M},..., \mu^2_{j/M},...,\mu^2_{M/M}
\end{equation}
The 16\,\% lower and upper values are discarded, and the new
extremes values of this vector give the limits of the 68\,\% confidence interval:
\begin{equation}
\mu^2_{\rm min} = \mu^2_{\ 0.16}\ ;\ \ \ \mu^2_{\rm max} = \mu^2_{\ 0.84}
\end{equation}
This is the probabilistic definition of 1\,$\sigma$ error bars, and we
therefore obtain the $\sigma_{+}$ and $\sigma_{-}$ asymmetric error bars through:
\begin{equation}
\sigma_{+} = \mu^2_{\rm max} - \overline{\mu^2}\ ;
\ \ \ \ \sigma_{-} =  \overline{\mu^2} - \mu^2_{\rm min}
\end{equation}
where $\overline{\mu^2}$ is the weighted average of the original sample $[\mu^2_i]$
using the weights $[w_i]$.
The same process can be applied using 2.5--97.5\,\% percentile limits to obtain the
error bars equivalent to $2\,\sigma$, and 0.5--99.5\,\% for $3\,\sigma$.

Alternatively, one can derive the bootstrapped variance $\sigma_{\rm BS}^2$
directly from the $\mu^2_k$ ensemble:
\begin{equation}
\sigma^2_{\rm BS} = \frac{1}{M} \sum_{j=1}^{M}\left(\mu^2_j-\overline{\mu^2}\right)^2
\end{equation}
The internal bias $b_{\rm BS}$ of the population is given by:
\begin{equation}
b_{\rm BS} = \overline{\mu_j^2} - \overline{\mu^2}
\end{equation}
This bias is naturally taken into account in the computation of the confidence
interval limits as described above.

\section{Statistical and systematic errors of the ratio of $\mu^2$ and $T^2$ \label{appendix_ratio}}

In the expression of $T^2$ of Eq.\,\ref{T2_eq}, we have to separate the contributions from the
systematic uncertainty on the calibrator knowledge, and the statistical error of the instrumental
measurement of $\mu^2$.
While these two terms contribute to the global uncertainty on the squared visibility $V^2$,
their nature is fundamentally different. While it is possible to reduce the statistical
error by averaging several measurements, the systematic uncertainty originating in
the calibrator diameter error bar will not be changed. This last term is therefore a
fundamental limitation to the absolute precision of the visibility measurement.
This limit can be reduced by using several calibrators,
or by selecting very small stars as calibrators. We then benefit from the fact that the visibility
function $V^2(B, \theta)$ for a stellar disk is nearly flat when the star is not significantly
resolved, and the resulting systematic error on $V^2$ remains small.

Considering a symmetric error bar on the assumed angular diameter of the calibrator,
the resulting error bar on the $V^2$ estimate is in general not exactly symmetric,
due to the non linearity of the visibility function.
In practice, asymmetric error bars are easily manageable numerically.
However, in order to simplify the notations in the present discussion,
we make the assumption that this asymmetry is negligible.

The estimation of the two kinds of error contributions relies on an approximation of the
Cauchy statistical law characteristics. When dividing two
normal statistical variables $x$ and $y$ of respective means and standard deviations
$(\overline{x}, \sigma_x^2)$ and $(\overline{y}, \sigma_y^2)$,
the resulting ratio $x / y$ follows a Cauchy statistics that has, strictly speaking,
no defined mean value.
It is therefore necessary to make an approximation for the case when
$\sigma_y \ll \overline{y}$. In this case, a second order approximation
of the mean $\overline{z}$ and variance $\sigma^2_z$ of $z = x / y$ is given by
Browne\,(\cite{browne02}):
\begin{equation}\label{eq_mu_z}
\overline{z} = \frac{\overline{x}}{\overline{y}} \left( 1 + \frac{\sigma_y^2}{\overline{y}^2} \right)
\end{equation}
\begin{equation}\label{eq_std_z}
\sigma_{z}^2 = \frac{\overline{x}^2}{\overline{y}^2} \left(\frac{\sigma_x^2}{\overline{x^2}}
+ \frac{\sigma_y^2}{\overline{y}^2} - \frac{\sigma_y^4}{\overline{y}^4}\right)
\end{equation}
It is not possible to obtain a meaningful average value of the ratio
$x/y$ if the standard deviation $\sigma_y$ of the denominator
$y$ is not small compared to its average value. As a remark, the average value of $x/y$
is in general different from $\overline{x}/\overline{y}$.

The average value of the transfer function $T^2$
and its associated statistical error bars are computed by
replacing in formulas \ref{eq_mu_z} and \ref{eq_std_z} the values of
$(\overline{x}, \sigma_x^2)$ and $(\overline{y}, \sigma_y^2)$ by the following terms:
\begin{equation}
\overline{x} \rightarrow \left[\mu^2\right]_{\rm \theta\,Cen}\ \ \ \ \ \ \ \ \ \ \ \ \ \ \ 
\sigma_x^2 \rightarrow \left[\sigma_{\rm stat} \right]_{\theta\,Cen}^2
\end{equation}
\begin{equation}
\overline{y} \rightarrow \left[V^2_{\rm theo}\right]_{\theta\,Cen}^2\ \ \ \ \ \ \ \ \ \ \ 
\sigma_y^2 \rightarrow 0
\end{equation}
Similarly, the systematic error is computed using the replacements:
\begin{equation}
\overline{x} \rightarrow \left[\mu^2\right]_{\rm \theta\,Cen}\ \ \ \ \ \ \ \ \ \ \ \ \ \ \ 
\sigma_x^2 \rightarrow 0
\end{equation}
\begin{equation}
\overline{y} \rightarrow \left[V^2_{\rm theo}\right]_{\theta\,Cen}^2\ \ \ \ \ \ \ \ \ \ \ 
\sigma_y^2 \rightarrow \left[\sigma_{\rm theo} \right]_{\theta\,Cen}^2
\end{equation}
Applying this computation to the numerical values found for $\theta$\,Cen, we find:
\begin{equation}
T^2 = 0.5009\ [\pm 0.0027]_{\rm stat}\ [\pm 0.0077]_{\rm syst}
\end{equation}
The uncertainty on this value is dominated by the systematic error.
The only remaining calibration step is now to divide the $\mu^2$ value obtained on
$\alpha$\,Cen\,A by the $T^2$ value. Again, we have to separate the two contributions on the
error by replacing in the above formulas the mean values and
standard deviations of $x$ and $y$ by the following terms for the statistical error:
\begin{equation}
\overline{x} \rightarrow \left[\mu^2\right]_{\rm \alpha\,Cen}\ \ \ \ \ \ \ \ \ \ \ \ \ \ \ 
\sigma_x^2 \rightarrow \left[ \sigma_{\rm stat} \right]_{\alpha\,Cen}^2
\end{equation}
\begin{equation}
\overline{y} \rightarrow T^2 \ \ \ \ \ \ \ \ \ \ \ \ \ \ \ \ \ \ \ \ \ \ \ \ 
\sigma_y^2 \rightarrow \left[\sigma_{\rm stat}\right]^2_{T^2}
\end{equation}
while the systematic error is computed using
\begin{equation}
\overline{x} \rightarrow \left[\mu^2\right]_{\rm \alpha\,Cen}\ \ \ \ \ \ \ \ \ \ \ \ \ \ \ 
\sigma_x^2 \rightarrow 0
\end{equation}
\begin{equation}
\overline{y} \rightarrow T^2 \ \ \ \ \ \ \ \ \ \ \ \ \ \ \ \ \ \ \ \ \ \ \ \ 
\sigma_y^2 \rightarrow \left[\sigma_{\rm syst}\right]^2_{T^2}
\end{equation}
We obtain the calibrated squared visibility of $\alpha$\,Cen\,A:
\begin{equation}
V_{\rm \alpha\,Cen}^2 = 0.00635\ [\pm 0.00033]_{\rm stat}\ [\pm 0.00010]_{\rm syst}
\end{equation}


\begin{acknowledgements}
D.S. acknowledges the support of the Swiss FNRS.
We wish to thank Dr. Guy Perrin for important comments
that led to improvements of this paper.
The interferometric data presented in this paper
have been obtained using the Very Large Telescope
Interferometer, operated by the European Southern Observatory at Cerro Paranal, Chile.
It has been retrieved from the ESO/ST-ECF Archive Facility (Garching, Germany).
Observations with the VLTI are only made possible through the efforts of the VLTI team,
to whom we are grateful.
\end{acknowledgements}

{}

\end{document}